\pdfoutput=1
\documentclass[journal=jctcce,manuscript=article]{achemso}

\usepackage{graphicx}
\usepackage{mathtools}
\usepackage{color}
\usepackage{amsmath} %helps with some of the multi-line math
\usepackage[caption=false]{subfig}
\usepackage{float} %Needed to make subfig work right
\usepackage{natbib}
\usepackage{multirow}
\usepackage{multicol}
\usepackage{pdflscape}
\newcommand{\plam}{\lambda}
\newcommand{\plamn}{\lambda_{n}}

\newcommand{\hnln}{h_{n}(\plamn)}

\newcommand{\bx}{\mathbf{x}}

\newcommand{\vecr}{r}

\newcommand{\eij}{\ensuremath{\epsilon_{ij}}}
\newcommand{\sij}{\ensuremath{\sigma_{ij}}}
\newcommand{\eii}{\ensuremath{\epsilon_{ii}}}
\newcommand{\sii}{\ensuremath{\sigma_{ii}}}

\newcommand{\sjj}{\ensuremath{\sigma_{jj}}}
\newcommand{\llangle}{\ensuremath{\left\langle}}
\newcommand{\rrangle}{\ensuremath{\right\rangle}}
\newcommand{\expect}[1]{\ensuremath{\llangle #1 \rrangle}}

\newcommand{\kcalmol}{\ensuremath{\text{kcal}/\text{mol}}}

\newcommand{\kcalmolK}{\ensuremath{\text{kcal}/\left(\text{mol}\cdot K\right)}}
\newcommand{\kjmol}{\ensuremath{\text{kJ}/\text{mol}}}
\newcommand{\una}{\ensuremath{u_{\text{unaffected}}}}
\newcommand{\ml}{\ensuremath{\ell}}
\newcommand{\ddx}[1]{\ensuremath{\Delta\left(\Delta #1\right)}}
\title{Rapid Computation of Thermodynamic Properties Over Multidimensional Nonbonded Parameter Spaces using Adaptive Multistate Reweighting}
\author{Levi N. Naden}
\affiliation{Department of Chemical Engineering, University of Virginia, Charlottesville, Virginia 22902, USA}
\author{Michael R. Shirts}
\email{michael.shirts@virginia.edu}
\affiliation{Department of Chemical Engineering, University of Virginia, Charlottesville, Virginia 22902, USA}

\begin{document}

\begin{abstract}
We show how thermodynamic properties of molecular models can be computed over a large, multidimensional parameter space 
by combining multistate reweighting analysis with a linear basis 
function approach. This approach reduces the computational cost to estimate 
thermodynamic properties from molecular simulations for over 130,000 tested parameter combinations from over a 
thousand CPU years to tens of CPU days.
This speed increase is achieved primarily by computing the potential 
energy as a linear combination of basis functions, computed from either modified simulation 
code or as the difference of energy between two reference states, which can be done without any simulation 
code modification. The thermodynamic properties are then estimated with the Multistate Bennett Acceptance Ratio (MBAR)  
as a function of multiple model parameters without the need to define \textit{a priori} how the 
states are connected by a pathway. Instead, we adaptively sample a set of 
points in parameter space to create mutual configuration space overlap. 
The existence of regions of poor configuration space overlap are 
detected by analyzing the eigenvalues of the sampled states' overlap matrix. The configuration space overlap to sampled 
states is monitored alongside the mean and maximum uncertainty to determine 
convergence, as neither the uncertainty or the configuration space overlap alone is a sufficient metric of convergence.

This adaptive sampling scheme is 
demonstrated by estimating with high precision the solvation free energies of charged particles of Lennard-Jones plus 
Coulomb functional form with charges between -2 and +2 and generally physical values of 
$\sij$ and $\eij$ in TIP3P water.  
We also compute entropy, enthalpy, and radial distribution functions of
arbitrary unsampled parameter combinations using only the data from
these sampled states and use the estimates of free energies over the
entire space to examine the deviation of atomistic simulations from
the Born approximation to the solvation free energy.

\end{abstract}
\maketitle

\section{Introduction}

Many applications of molecular simulations require searching over large parameter spaces 
to predict or match physical observables. Molecular simulation parameters such as charges, Lennard-Jones dispersion and repulsion parameters,  
as well as bonds, angles, and torsion force constants determine the energies and 
probabilities of configurations in simulations, and thus in turn, determine 
what thermodynamic properties will be observed. The ability to accurately estimate
thermodynamic properties 
without the need for laboratory experiments has the potential to save both time and resources in fields 
such as polymer~\cite{Wang2014} and solvent~\cite{Harini2013} design as well as drug 
discovery.~\cite{BaronEd2012,Li2012c,Hansen2014} This time and cost savings is important both in the design of new molecules, 
where properties are unknown, and 'reverse property prediction' where a model or molecule 
is designed to match specific experimental targets, such as designing metal-organic 
frameworks (MOFs) with specific gas loadings.~\cite{Yang2013b}

Assigning proper parameters in a molecular simulation given a set of experimental data 
becomes more difficult as the number of free parameters increases.  
Some experimental parameters, such as bond lengths and angles, are
relatively easy to estimate using small-molecule crystallographic structures and quantum
chemistry. However, nonbonded parameters, such as partial charges and
dispersion terms, are much more difficult to choose as they
are model parameters that do not directly correspond to laboratory
observables such as bond length. Instead, possible nonbonded parameters are constrained by sets
of experimental observables such as transfer free energies, heats of
vaporization, densities, and heat capacities.

Identifying nonbonded model parameters consistent with a set of
experimental thermodynamic values requires expensive iterative,
self-consistent simulations or even more expensive gradient
optimizations.~\cite{Monticelli2013}
Most condensed phase force 
fields~\cite{Mackerell1998,Vanommeslaeghe2011,Oostenbrink2004,Jorgensen1996,Kaminski2001,Pearlman1995,Wang2004} 
are parameterized by manual iterative fitting to a training set of experimental 
thermodynamic data for a small set of molecules chosen to represent a broader spectrum of similar molecules.~\cite{VanGunsteren1998,VanGunsteren2006} 
Accurate fits are required to predict properties of biological systems or complex mixtures 
where group contribution methods such as UNIQUAC~\cite{Abrams1975} 
and UNIFAC~\cite{Fredenslund1975} are inadequate.

Some of the most computationally expensive properties to estimate involve the 
free energy differences between two states, such as the solvation free energy, which 
is the free energy difference of two systems as one solute molecule moves from solution to 
vapor, or activity coefficients, which measure the deviation of the chemical potential of a species from ideality. 
Accurately computing free energy differences (or equivalently, chemical potential differences) 
also provide a way to compute many other thermodynamic properties as they can be derived from the derivatives of the free 
energy with respect to temperature ($T$), pressure ($P$), volume ($V$), and number of 
particles ($N_{i}$).

Estimating free energy differences between two thermodynamic states accurately requires
designing a thermodynamic path between the states. Paths that are both 
computationally and statistically efficient, especially for full deletion or insertion of 
a molecule into a dense fluid, are nontrivial to design and must often include a number of 
non-obvious, nonphysical intermediates.~\cite{Frenkel2002,Pitera2002,Blondel2004,Crooks2007,Steinbrecher2007,Hritz2008,Riniker2011b,Buelens2011,Pham2011,Pham2012,Naden2014,Naden2014b,Shirts2003,Steinbrecher2011,Mobley2009} 
The end states and multiple intermediate states along the thermodynamic path must be 
sampled to create good configuration space overlap between the end states, which is required to accurately 
estimate free energy differences between the endpoints.~\cite{Weinhold1975,Bennett1976,Frenkel2002,Boresch2003,Wu2005,Hansen2014} 
Technically, we require good overlap in the full phase space, both configurations 
and velocity, but because velocities are thermalized in essentially all systems 
of thermodynamic interest, it is the configuration space that we must generally worry about when connecting states together.

If one wishes to compute free energies of solvation for many different
parameterizations of the same molecule, the direct approach of calculating an entire thermodynamic 
pathway for each parameter choice is extremely expensive for highly accuracy calculations. 
Additionally, examining the differences
in free energies due to a small change of parameters is particularly
difficult because we must take the differences between two similar numbers
with independent statistical error. 

Reweighting methods can help solve both the problem of expense and the 
problem of cancellations of errors. In a recent study, our group showed
how multistate reweighting can directly calculate the $\Delta \Delta
G$ between two different long-range interaction approaches, with very
small uncertainties for relatively low computational cost.~\cite{Paliwal2013b} 
The expense is lowered by constructing thermodynamic cycles directly connecting Hamiltonians 
with similar parameters and thus significantly overlapping configurational spaces. 
These small uncertainties are
possible because MBAR can directly calculate the covariances between
the two free energies through analyzing all potential energy differences, rather than only 
uncorrelated calculations. This same approach can be applied to small
changes in parameters.

Estimating properties with reweighting methods requires constructing a thermodynamic path between 
the different parameterizations.  It also requires potentially significant computational resources to 
perform simulations with parameters that have properly overlapping configurations. 
The combination of these two requirements adds theoretical and practical 
limitations to simultaneously searching large, multidimensional nonbonded parameter spaces.
\begin{enumerate}
 \item The space of nonbonded parameters is often at least multiple dimensions \textit{per particle} or particle type.  
  For example, the nonbonded parameters of charge ($q$) and least two Lennard-Jones-like terms ($\eij$, $\sij$) can result in at least three parameter dimensions per particle type. 
 \item There is no obvious way to define computationally efficient thermodynamic paths between any two points in these multidimensional spaces or select {\em a prior} simulation 
  points in this space that give rise to low error estimates of thermodynamic 
  properties across the entire space.
 \item Reweighting methods requires computing energies from the sampled configurations to other 
  sampled states, and any unsampled state of interest. 
  This re-computation typically requires re-running the simulation force loops 
  over all generated configurations for each combination of parameters of interest. 
  The computational cost to search such a multidimensional space of nonbonded parameters 
  scales, at best, linearly with the number of samples, and at worst 
  quadratically with the number parameter combinations, since data may be collected 
  with simulations at each parameter combination.~\cite{Naden2014}
\end{enumerate}

\begin{figure}[H]
  \centering
  \includegraphics[width=0.55\textwidth]{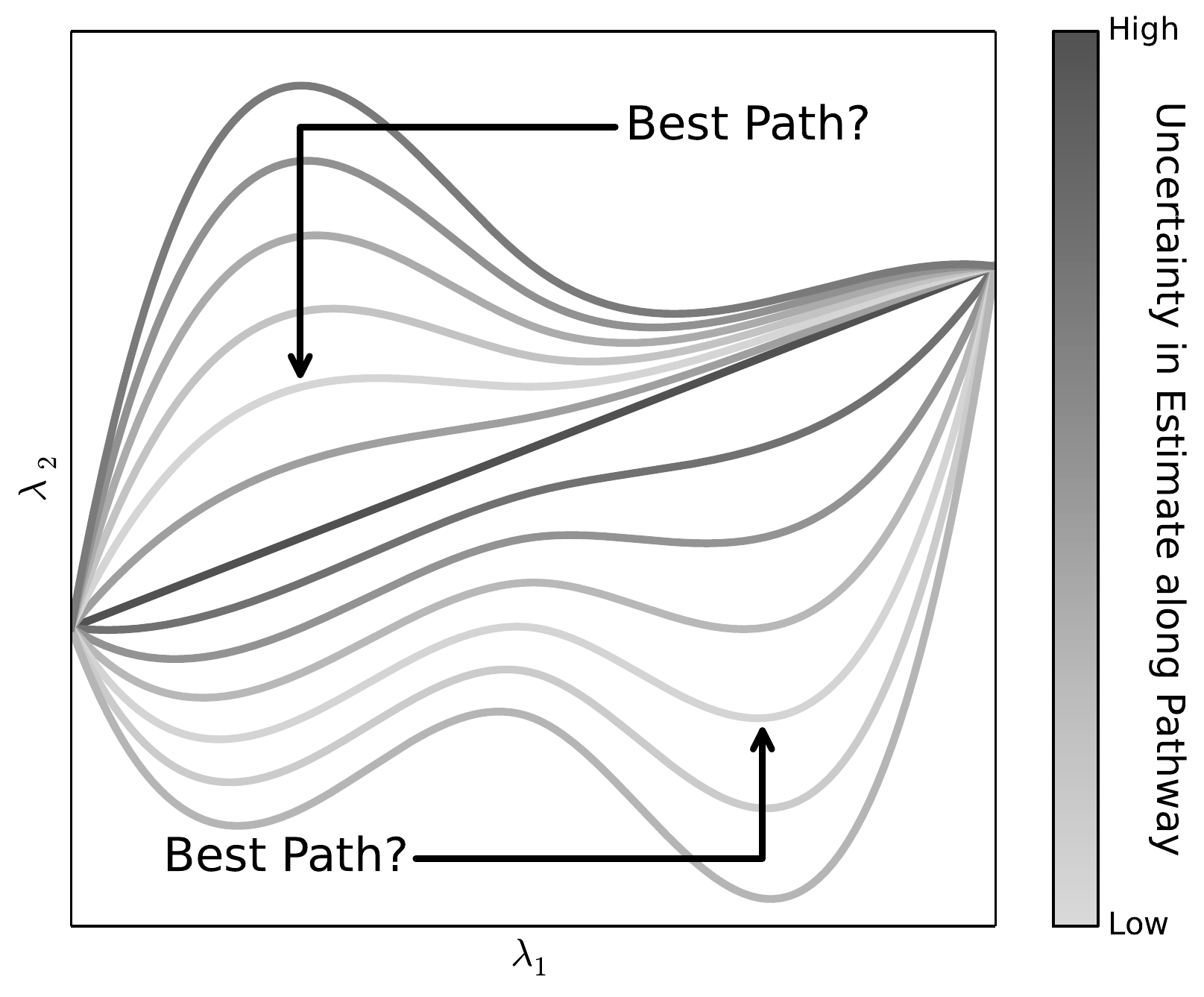}\\  
  \caption[]{{\bf Defining the ``best'' thermodynamic path between arbitrary states 
  is a nontrivial problem.} This figure shows several multidimensional thermodynamic paths connecting 
  two states, and the relative uncertainty of each pathway shown by a gray scale gradient 
  of each curve. The path with shortest Euclidean distance in parameter space ($\plam_{1}$, $\plam_{2}$) has a large uncertainty, 
  and at least two of the other paths have lower uncertainty. However, it is unclear if sampling 
  a single ``best path,'' or a combination of multiple low variance paths will have the highest 
  computational efficiency for a target statistical error. The most computationally 
  efficient sampling scheme may not be along any single path at all, and instead may be achieved by sampling 
  a non-parameterized states in a multidimensional parameter space.}
  \label{fig:ExamplePath}
\end{figure}

Designing efficient thermodynamic paths through arbitrary thermodynamic states is a challenging task,~\cite{Pham2011,Naden2014,Naden2014b}
but designing paths in multidimensional parameter spaces adds additional complexities. 
An example in Fig.~\ref{fig:ExamplePath} demonstrates the challenges of identifying low-uncertainty paths in multiple dimensions. 
This figure shows two arbitrarily defined thermodynamic states in a two-dimensional parameter space
and attempts to draw pathways between them with high mutual configuration space overlap, providing 
low error estimates. However, the choice of which path to sample to connect the states is not 
immediately obvious. The shortest Euclidean path in parameter space has large uncertainty, 
but two alternative paths have low uncertainty. This sort of multidimensional space raises questions 
with no obvious answers: How can we {\em a priori} identify which paths 
have more mutual overlap (i.e. result in simulations with lower uncertainty) without exhaustively sampling the system? 
Could samples drawn from both paths but with different proportions provide lower uncertainty than sampling either path 
by itself?

Previous research on identifying low-variance paths along a single
coupling parameter used local minimization of total variance along the
path.~\cite{Pham2011,Naden2014,Naden2014b}  However, in multiple
dimensions, since multiple potential low-variance paths could exist, local optimization is unlikely to 
identify the most efficient path between states of interest, nor whether it would be more optimal to traverse both paths. 
The identification of an optimal path choice is made more complicated if we are interested in
all the mutual free energy differences between multiple states, since we must identify 
a path or a network of paths connecting all of the states, with samples collected along each of these states.
One would ideally want ad hoc rules estimating the computational efficiency of 
these paths, determined \textit{a priori} to avoid unnecessary sampling. Defining such 
rules for a diverse set of chemical systems becomes increasingly complex as the dimensionality 
and the number of states increases. 
Removing the need to define these difficult multidimensional 
paths significantly lowers these barriers to searching through large parameter spaces for optimal parameters.

We can remove the need to explicitly define thermodynamic paths between 
states in multiple dimensions with \textit{multistate} reweighting methods. These methods, such as the 
Multistate Bennett Acceptance Ratio~(MBAR)~\cite{Shirts2008} and the Weighted Histogram 
Analysis Method~(WHAM),~\cite{Kumar1992,Kumar1995} allow 
analyzing samples from anywhere in the parameter space to compute properties anywhere else in 
the space, without the need to define which 
thermodynamic states are adjacent as is needed for the original Bennett Acceptance 
Ratio~(BAR).~\cite{Bennett1976} MBAR has the important advantage over WHAM in that sampled 
configurations do not need to be binned along a pre-defined thermodynamic path for the analysis
which is particularly important in multidimensional spaces, where it becomes 
increasingly difficult to populate histogram bins. 
Any given combination of nonbonded parameters will share a high degree of configuration space 
overlap with a large number of similar parameter sets, though it is not always clear 
{\em a priori} which ones those will be. MBAR takes into account all of the information 
from all sampled states in determining the free energy differences between any two states, 
sampled or unsampled. 
Therefore, with moderate sampling at the right parameter combinations, 
thermodynamic properties can be estimated at a large number of parameter combinations, 
even without explicit sampling of every parameter combination, so long as for each choice 
of parameters there is some degree of configuration space overlap with some combination of sampled parameters.

With the limitation of explicitly defining a thermodynamic path removed, we can focus on 
decreasing the computational cost of gathering the statistical information  
required for multistate reweighting calculations of thermodynamic information. Reweighting methods require the energy of each 
sampled configuration to be known at each state of interest to compute free energy differences and other properties between states.
Computing a configuration's energy multiple times with different energy functions 
is a time limiting step for calculating thermodynamic properties across large parameter spaces. 
Collecting the energy of each configuration at each state usually requires running 
the simulation code, or at least the inner force loop of simulation code, multiple times on each configuration. 
If we only are interested in reweighting the properties from a single sampled state~\cite{Zwanzig1954}, we 
would only have to carry out this computation once 
per configuration ($N$) generated at a sampled state ($K_{s}$), per unsampled state of 
interest ($K_{u}$). This results in 
energy calculations that scale as $\mathcal{O}\left(N(K_{u}+K_{s})\right)$, assuming equal 
sampling per sampled state.
However, such estimates are known to be
statistically inefficient and prone to substantial bias when overlap is not substantial.~\cite{Shirts2005a,Paliwal2011,DeRuiter2012a} 
Multistate methods, like MBAR,~\cite{Shirts2008} require calculating the energy for 
each configuration at each sampled state as well as each state of interest, so the scaling is
quadratic in the number of sampled states as $\mathcal{O}\left(N(K_{u}+K_{s}^{2})\right)$. 
This is not a burdensome task for a small 
number of states along a single thermodynamic path, making it well worth using 
multistate simulations regardless of how expensive recalculating the energies of this configurations are.
 However, as tens or hundreds of thousands of states of 
interest and hundreds of sampled states are considered, this scaling becomes a computational bottleneck that must be overcome.

We can reduce the cost to compute energies at thermodynamic states to computationally 
trivial vector multiplication by defining the energies using linear basis 
functions.~\cite{Buelens2011,Naden2014,Naden2014b}
Energies calculated using a basis function approach can be most generally written by
\begin{equation}\label{eq:basis}
 u(\vecr,\lambda) = \sum_i^n h_i(\plam)u_i(\vecr) + u_{\text{unaffected}}(\vecr)
\end{equation}
where $u(\vecr,\lambda) = \beta U(\vecr,\lambda)$ is the reduced energy as a function of both the configuration 
$\vecr$ and some (possibly multidimensional) alchemical coupling parameter 
$\lambda$; $h_i(\plam)$ are a set of nonphysical, alchemical switches that are 
independent of configuration; $u_i(\vecr)$ are the 
basis functions; $n$ is the total number of basis 
function and alchemical switch pairs; and $u_{\text{unaffected}}(\vecr)$ is the system's 
potential energy not dependent on the alchemical variables. 
This approach computes the energy of a configuration at 
any thermodynamic state by scalar multiplication of the configuration dependent 
basis functions, which only have to be computed once per configuration. 
The vector multiplication can eliminate the need to run the inner force loop on a 
configuration more than once, reducing the computational cost of 
evaluating energies from  
$\mathcal{O}\left(N(K_{u}+K_{s}^{2})\right)$, to $\mathcal{O}\left(NK_{s}\right)$, which is 
simply the total number of sampled configurations.
The alchemical switches can take any form, 
so long as $\vecr$ remains part of the basis functions and not the switches. This form of the potential energy is in contrast to forms such as the 
soft core form of repulsive interactions~\cite{Beutler1994,Zacharias1994}, which cannot be 
represented as sums of separable combinations of $\vecr$ and $\lambda$.

In this study we combine multistate reweighting methods with a linear basis 
function approach to compute 
thermodynamic properties over a large nonbonded parameter space. To demonstrate the process, we look 
at the Lennard-Jones parameters $\eii$ and $\sii$, and partial charge, $q_{i}$, 
for a single particle in explicit solvent. This approach could can aid in future 
large parameter space searches to quickly find a range of nonbonded parameters and fine tune a 
fitting or optimization procedure. The relative free energy, enthalpy, and entropy of solvation are explored 
as these are some of the most computationally expensive properties to estimate. 
We also estimate the Born solvation free energy of charging, compare our
results to specific ion free energies computed from others, and compute radial 
distribution functions. The techniques shown here are generalizable to other thermodynamic properties.

We explore both a two-dimensional and a three-dimensional parameter space. A 2-D 
parameter space in $\eii$ and $\sii$ is first explored as a test of searching through parameter space. 
A larger, 3-D parameter space in $\eij$, $\sii$, and $q_{i}$ is then also explored and iterative 
simulations are carried out to reduce the statistical error in the estimate of solvation 
properties across the entire range of parameters.

%%%%%%%%%%%%%%%%%%%%%%%%%%%%%%%%%%%%%%%%%%%%%%%%%%%%%%%%%%%%%%%%%%%%%%%%%%%%%%%%%%%%%%%%%%
\section{Theory \label{sec:theory}}

The notation in this study is as follows. $\sii$ and $\eii$ are the Lennard-Jones parameters, 
and $q$ is the charge of a particle. $\epsilon_{0}$ is the permittivity of vacuum. $u(\vecr)$ are basis functions of the potential energy function 
and $h(\plam)$ are the alchemical switches. Subscripts $E$, $R$, and $A$ 
denote an electrostatic, Lennard-Jones repulsive, and Lennard-Jones attractive term 
respectively, e.g. $u_E(\vecr)$ is the electrostatic basis function. The subscripts $i$, 
$j$, and $k$ on nonbonded parameters denote arbitrary atoms, 
and subscripts $X$, $Y$, and $Z$ denote an explicit set of parameters which define a 
thermodynamic state. 
and have fixed values but not explicitly defined to be general. 
Subscript $S$ denotes a solvent particle. 
The subscript $\ml\,$ will be used for summation indices, and $C$ represents a collection of constants. 

\subsection{Representing nonbonded parameter space with basis functions}

We generalize the potential energy to simplify writing the energy of any 
potential in multidimensional space. The three nonbonded parameters explored here 
lead to a pairwise nonbonded 
potential energy between two point particles a distance $r$ apart
\begin{equation}
 u(r) = \frac{4\eij\sij^{12}}{\vecr^{12}} + \frac{-4\eij\sij^{6}}{\vecr^{6}} + \frac{q_{i} q_{j}}{4 \pi \epsilon_{0} \vecr}\text{.}
 \label{eq:pairwise}
\end{equation} 
Eq.~\eqref{eq:pairwise} can be more generally written as 
\begin{alignat}{2}\label{eq:gen-pair}
 u(r) &= \frac{C_{12}}{\vecr^{12}} + \frac{C_6}{\vecr^{6}} + \frac{C_1}{\vecr} \\
      & = \sum_{\ml} \left(\frac{C_{n}}{r^n}\right)_{\ml}
\end{alignat}
where $n$ takes discrete values of 12, 6, or 1 depending on the index of $\ml$ and each $C_{n}$ 
corresponds to the power of $r^{-n}$. The energy of a configuration at any point in parameter space 
is found by adding an alchemical switch, $\hnln$, to each term of Eq.~\eqref{eq:gen-pair}. 
Each $\plamn$ can vary independently each other, allowing a multidimensional representation 
of the energy in terms of the parameters.
The alchemical switches scale each of the $12$, $6$, and $1$ terms to produce each of the target thermodynamic states. 
The total potential is then
\begin{equation}\label{eq:gen-alch-pair}
 u(\vecr,\plam) = \una(\vecr) + \sum_{\ml} \left(\frac{\hnln C_{n}}{r^n}\right)_{\ml}\text{.}
\end{equation}

Computing the basis functions can be done either directly in code or in post-processing with 
fixed reference states. The most computationally efficient way to compute the basis functions 
would be to have the simulation package provide them at run time. However,
most simulation packages will not allow the user direct access to the basis function values 
without heavily modifying its code, since usually only the 
total potential energy or the total $\lambda$-dependent energy is required. 
An alternative solution which avoids any code modification is to choose 
two fixed 
reference states and compute the basis functions as a difference in energy, as was done in this study and explored below. 
This alternative approach means we must 
run the force calculations at least three times for any sampled configurations: once while the samples 
are generated, and once for each reference states. However, the computational 
cost is only $\mathcal{O}\left(3NK_{s}\right)$, which still scales much
better than $\mathcal{O}\left(N(K_{u}+K_{s}^{2})\right)$ as $K_{u}$ and $K_{s}$ increase.

The potential can be represented as linear combination of
alchemical perturbations around a fixed reference particle at state $X$ with
respect to a second reference particle at state $Y$ as
\begin{alignat}{2}
 u(\vecr,\plam) &= \una(\vecr) + \sum_{\ml} \left[\frac{(1-\hnln)C_{n,X}(r) + \hnln C_{n,Y}(r)}{\vecr^{n}}\right]_{\ml} \notag \\
                &= \una(\vecr) + u_{X}(\vecr) + \sum_{\ml} \left[\frac{\hnln\Delta C_{n,XY}(r)}{\vecr^{n}}\right]_{\ml} \label{eq:xform-total-delta}
\end{alignat}
where $\Delta C_{n,XY}(r) = C_{n,Y}(r) - C_{n,X}(r)$ and $u_{X}(\vecr)$ is the complete nonbonded pairwise 
potential for particle $X$ alone.
The computed basis functions are then calculated as the energy difference between the two 
reference particles, and the unmodified potential energy of particle $X$ becomes part of $\una(\vecr)$. 
The potential energy at arbitrary state $Z$ can now be 
computed using this perturbation. This reference state approach 
makes computing the basis functions possible without major simulation code changes. 
The numerical error should be monitored for any round off error since two similar 
energies are subtracted, as we discuss in Section~\ref{sec:numerics}.

Unlike with standard alchemical transformations between $\plam=0$ and $\plam=1$, 
the accessible parameter space is not bounded by the reference states. Consider an 
arbitrary state, $Z$, with parameters outside the range of the parameters $C_{n,X}$ and $C_{n,Y}$. 
The values of alchemical switches defining $Z$ would then fall outside the standard $[0,1]$ domain. States 
which fall outside this domain still have physical meaning in this context, unlike 
states with $\plam$ outside $[0,1]$ have no meaning for particle insertion or deletion
simulations. For example, the expanded domains in our previous studies~\cite{Naden2014,Naden2014b} 
served no practical purpose since $\hnln < 0$ 
represented a state where particles had an attractive atomic center, and $\hnln > 1$ represented a 
state ``more than fully coupled.'' 

The number of terms in Eq.~\eqref{eq:xform-total-delta} will increase quadratically as the number of 
interaction sites in the solute increase, increasing the number of $\eij$ and $\sij$ terms. 
However, geometric mixing rules can avoid such a large increase in terms. 
Details of how geometric mixing rules allow $\hnln$ terms to be solvent-independent
% are given in supplementary material in section~\ref{sec:solvent}. %with xr
are given in supplementary material in section~S.1.

%%%%%%%%%%%%%%%%%%%%%%%%%%%%%%%%%%%%%%%%%%%%%%%%%%%%%%%%%%%%%%%%%%%%%%%%%%%%%%%%%%%%%%%%%%
\section{Experimental Design \label{sec:exp_design}}

Molecular dynamics (MD) simulations of a single particle in explicit TIP3P water were carried out with 
GROMACS~4.6.5~\cite{Pronk2013,GromacsWeb} compiled in double precision. 
NVT equilibration was carried out for $100\,\text{ps}$, followed by NPT equilibration for 
$500\,\text{ps}$, followed by NPT production simulations of $6\,\text{ns}$ per simulated parameter 
combination. 
Temperature was held at $298\,\text{K}$ and coupled through Langevin dynamics with a time constant of 
$5\,\text{ps}$. Pressure (for NPT simulations) was held at $1\,\text{atm}$ and coupled with a 
Parrinello-Rahman barostat,~\cite{Parrinello1981,Nose1983} with a time constant of $5\,\text{ps}$, 
and a compressibility of $4.5\cdot10^{-5}$ bar$^{-1}$.

Solvation properties were estimated over a grid of nonbonded parameters for the particle.
For the 2-D case, the parameter ranges are
$0.0239\,\kcalmol \le \eii \le 0.8604\,\kcalmol$ ($0.1\,\kjmol \le \eii \le 3.6\,\kjmol$) and
$0.25\,\text{nm} \le \sii \le 1.2\,\text{nm}$. This range was chosen to include the 
largest possible particles in the OPLS-AA force field.~\cite{Jorgensen1996,Kaminski2001} 
with additional parameters to test the limits of the reweighting methods.
Solvation properties were calculated on a 
square grid of $\eii$ and $\sii$ with 151 grid points in each dimension for 22,801 total 
parameter combinations.
Initial grid points were distributed uniformly in $\eii$ and uniformly in $\sii^3$ so that sampling 
was done 
approximately proportional to the free energy of cavitation.~\cite{Huang2001,Grigoriev2007}
Relative solvation properties were computed from 
the reference parameters $\eii=0.1816\,\kcalmol$, $\sii=1.0170\,\text{nm}$ so the reference 
was roughly in the middle of the $\sii^{3}$ space.

For the 3-D case, the parameter ranges are $0.0239\,\kcalmol \le \eii \le 0.8604\,\kcalmol$, 
$0.25\,\text{nm} \le \sij \le 0.958\,\text{nm}$, and $-2.0 \le q_i \le +2.0$ 
in units of elementary charge with each 
dimension having 51 points for $51^3$ grid points in the cube and a total of 132,651 
parameter combinations. To improve resolution in some of the images, 101 $\sii$ states 
were estimated for $101 \cdot 51^2$ grid points and 262,701 combinations. The reference 
state chosen for this set test was $\eii = 0.0502\,\kcalmol$, $\sii=0.5732\,\text{nm}$, and 
$q_{i} = 0.0$. 
This set covers particles in the OPLS-AA force field from hydrogen (bound to a carbon), through 
the largest ions. The 
reference state was chosen to show how properties with low uncertainty can be estimated to 
particles of very different sizes and charges through iteratively selecting new parameter 
combinations to simulate. The spacing for initial sampling for $\eii$ and $\sii$ remains 
unchanged from the 2-D case, and the sampling in $q_i$ was done proportional 
to $q_{i}^{2}$ in keeping with Born theory for the free energy of solvation of charged spheres.
This choices resulted in initial sampling at charges $\pm 2.0000$, $\pm 1.8516$, $\pm 1.6903$, 
$\pm 1.5119$, $\pm 1.3093$, $\pm 1.0690$, $\pm 0.7559$, and $0.0000$, all with the reference state choices of $\eii$ and $\sii$.
Starting molecular geometries were generated with AMBERTOOLS's LEaP~\cite{amber12} and 
initial equilibration was carried out with the reference state parameters. All other 
solutes started their equilibration process from the final frame of the reference ion's NPT 
equilibration step.

Details about specific algorithm to compute basis functions with GROMACS and all input 
files are included in the supplementary information for this article.~\cite{NadenSupNote} 
The analysis code can be found on GitHub.~\cite{NadenGithub}
%LIST OF FILES TO INCLUDE, this can be in the repo.
%nvt.mdp
%npt.mpd
%md.mdp
%ljspheres_esq.top - completed version
%ljspheres_esq.gro - initial, pre equilibration
%npt6.gro - post ion 6 equilibration
%Amber input files? I did a manual renaming of atoms with a find and replace. If yes: *.leap.in, leaprc.*
%Instructions on which reference states to use to compute the basis functions in GROMACS:
%%Full at A
%%Full at B
%%Repulsive at A
%%Repulsive at B
%%Null (run trajectory with solvent absent, gives u_unaffected
%%Charge +1 (basis function)
%%Charge +0.5 (long range, interactions with periodic copies)

%%%%%%%%%%%%%%%%%%%%%%%%%%%%%%%%%%%%%%%%%%%%%%%%%%%%%%%%%%%%%%%%%%%%%%%%%%%%%%%%%%%%%%%%%%
\section{Results and Discussion}

\subsection{Solvation properties over a 2-D parameter space}

\begin{figure}[H]
  \centering
%   \subfloat[Original 11 sampled states]{\label{fig:LJ11}\includegraphics{images/smallLJParameterization_nstates11.png} }~
%   \subfloat[Naive additional sampling at 12 states and spot checks of Lennard-Jones spheres]{\label{fig:LJ12}\includegraphics{images/smallLJParameterization_nstates12.png} }\\
  \subfloat[Original 11 sampled states]{\label{fig:LJ11}\includegraphics{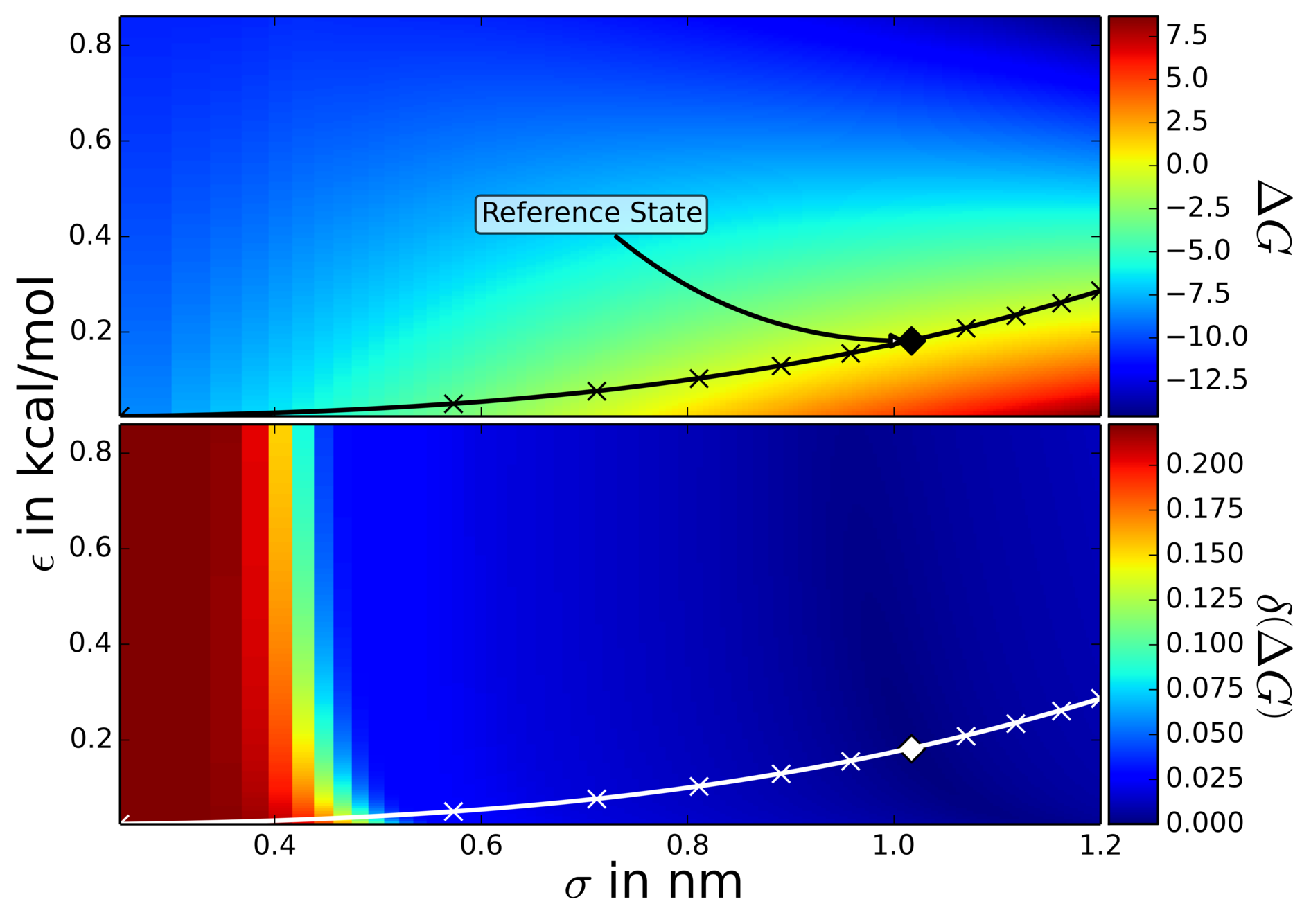} }~
  \subfloat[Naive additional sampling at 12 states and spot checks of Lennard-Jones spheres]{\label{fig:LJ12}\includegraphics{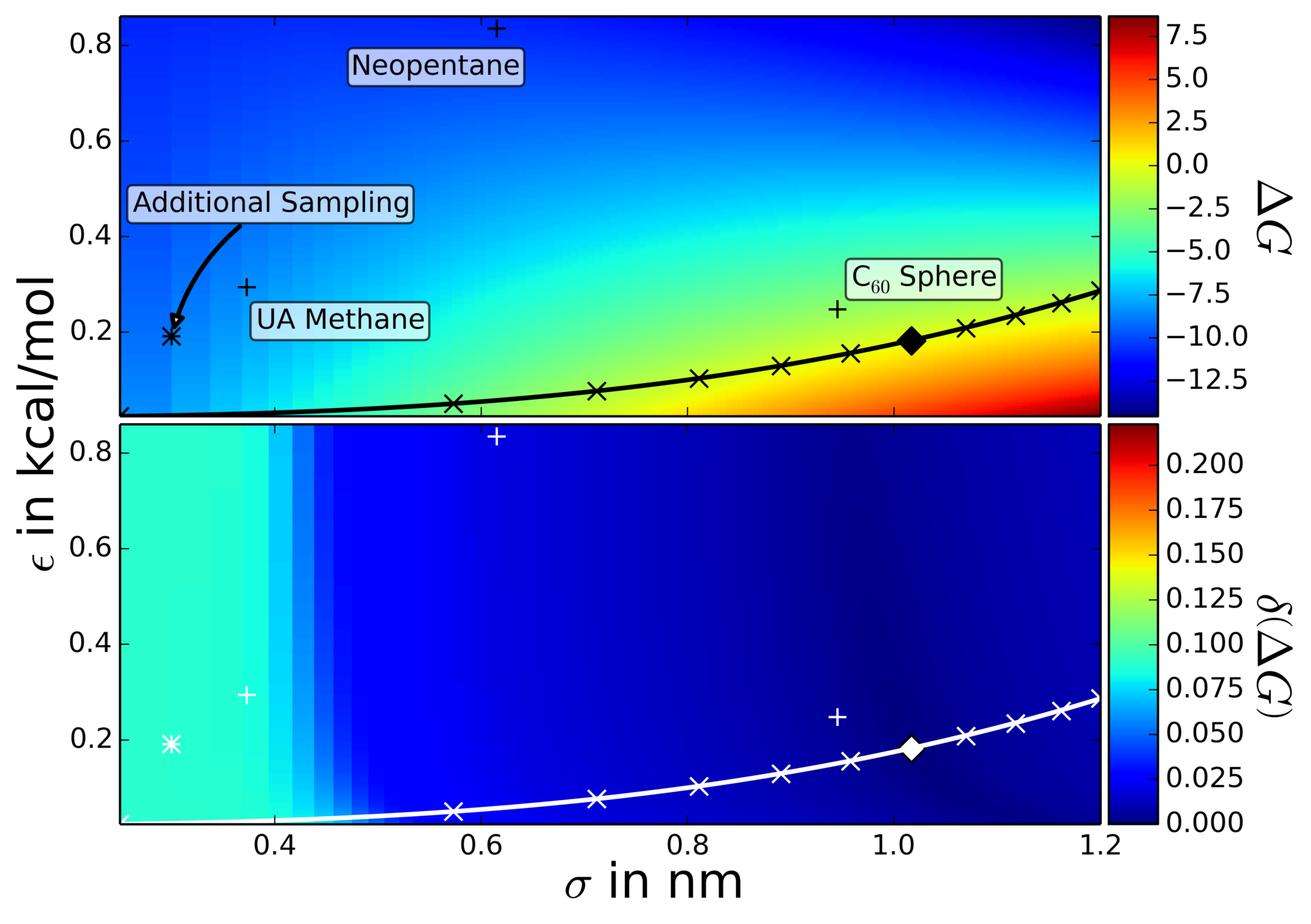} }\\
  \caption[]{{\bf The free energies at any combination of nonbonded parameters can be 
  predicted in post-processing and regions of large uncertainty can be quickly found}. 
  Shown is the free energy (top panels) and statistical error in the free energy 
  (bottom panels) for $151^2$ parameter combinations of $\eii$ and $\sii$ with no charge. 
  Samples were drawn at locations shown by an X and the drawn line is to guide the eye. 
  The free energies are all relative to the reference state shown by the diamond. \subref{fig:LJ11} 
  shows the estimates drawing samples from 11 states only. A large region of uncertainty 
  is seen at small $\sii$. \subref{fig:LJ12} A 12$^\text{th}$ state was sampled in the 
  region of high uncertainty, reducing the uncertainty in the whole region. The relative 
  free energy of solvation was compared to chemically realistic spheres through direct 
  simulation and is within error of the $\delta\Delta G$.
  Free energy is shown in units of $\kcalmol$.}
  \label{fig:LJOnlySampling}
\end{figure}

With the combination of methods described above, we can efficiently and accurately calculate the free energy of solvation and 
other thermodynamic properties over multidimensional parameter spaces. Figure~\ref{fig:LJOnlySampling} 
shows the free energy, and error in free energy of uncharged Lennard-Jones spheres evaluated at $151^2$ 
combinations of $\eii$ and $\sii$.  The free energy differences were 
estimated using MBAR implemented in the {\tt pymbar} package.~\cite{Shirts2008} 
One of the main keys to making this calculation feasible is that the linear basis function 
approach allows rapidly calculation of potential energies in post-processing. 
Reconstructing the potential energies required for free energy estimates through vector 
operations takes only seconds on a single core of a desktop computer's CPU.  
The same evaluation of energies would have to be run through the inner force loops 
at all $151^2$ states without the linear basis function method, scaling as 
$\mathcal{O}\left(N(K_{u}+K_{s}^{2})\right)$. We ran each sampled state's trajectory 
through 
single point energy calculations with GROMACS
estimating the potential under every other 
sampled state thermodynamic conditions to quantify the computational cost. 
Each simulation of 30000 samples took over 1500 CPU seconds to re-evaluate 
the energies of the given trajectory. For reference, the average time to run the simulation 
on the same hardware was 25 CPU hours. If we make a conservative calculation and 
assume each re-run of the inner force loop took 
the minimum 1500 CPU seconds, the 12 sampled states would have taken 13 CPU years 
to run each configuration through the $151^2$ parameter combinations. 
This computational cost illustrates the primary speed improvement over
re-running the inner force loop code, since time required to collect samples and estimate 
free energies is not affected by how the potential energies are computed in post-processing.

For the 2D problem, regions of large uncertainty can quickly identified by visual inspection, and 
additional samples can drawn to reduce uncertainty. Fig.~\ref{fig:LJ11} shows
the estimates of the solvation free energy when sampling from only 11 equivolume spaced states. We can 
see in the figure that parameter 
combinations with roughly $\sii < 0.5\,\text{nm}$ have high estimated uncertainty 
with respect to the reference state.
We can naively sample by a single additional state in this region which drastically reduce 
the error in our estimation across this range as shown in Fig.~\ref{fig:LJ12}. The error is a much steeper function of $\sii$ than 
$\eii$ in these ranges since large particles share virtually no configuration space 
overlap with small particles in a dense fluid due to changes in the packing of solvent particles around small solutes.

The linear basis functions approach reproduces the results from  
direct fixed-parameter solvation simulations. Fig.~\ref{fig:LJ12} is annotated to show 
where several chemically realistic Lennard-Jones spheres fall in the parameter 
space. Solvation simulations 
were run for the parameters of united atom (UA) 
methane,~\cite{Kaminski2001,Shirts2005a,Paliwal2011} neopentane~\cite{Kuharski1984}, and a 
sphere roughly the size of a C$_{60}$ molecule.~\cite{Naden2014} The relative free energies 
are statistically indistinguishable between the direct solvation simulations and those 
computed in Fig.~\ref{fig:LJ12}. The exact numbers and methods for the solvation 
simulations are shown in the supplementary material in 
% Section~\ref{sec:direct_solv}.%xr
Section~S.2.

\subsection{Solvation properties over a 3-D parameter space}

\begin{figure}[H]
  \centering
%   \subfloat[]{\label{fig:n21-f6} \includegraphics{images/smallSingleFrame_n21__f6.png} }~
%   \subfloat[]{\label{fig:n21-odd}\includegraphics{images/smallSingleFrame_n21__f5.png}}\\
%   \subfloat[]{\label{fig:n21-f31}\includegraphics{images/smallSingleFrame_n21__f31.png}}\\
  \subfloat[]{\label{fig:n21-f6} \includegraphics{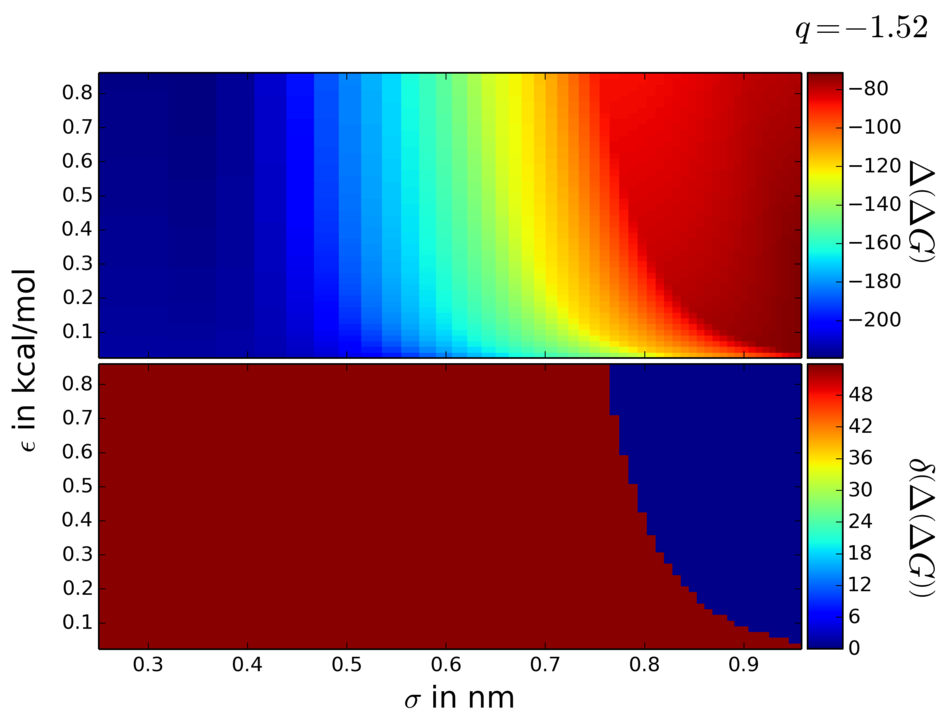} }~
  \subfloat[]{\label{fig:n21-odd}\includegraphics{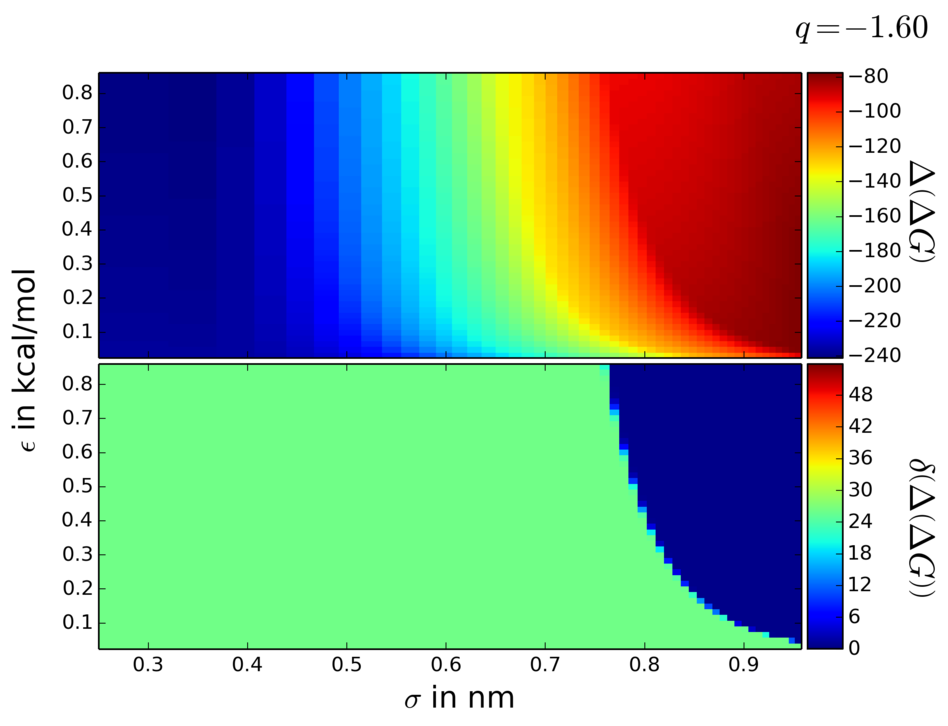}}\\
  \subfloat[]{\label{fig:n21-f31}\includegraphics{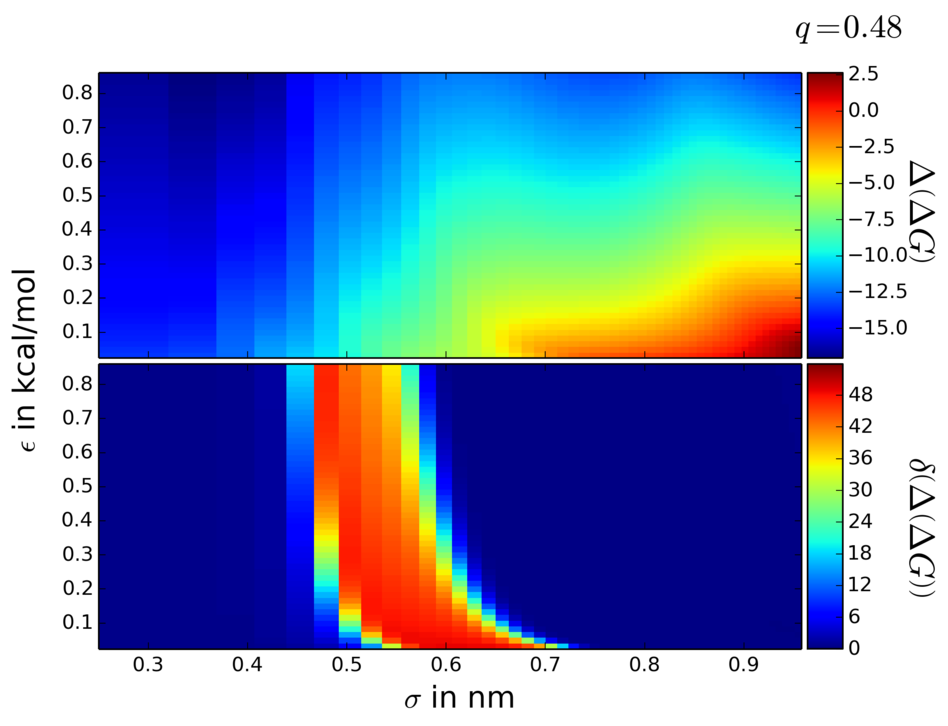}}\\
  \caption[]{{\bf The free energies in a multidimensional nonbonded parameter space can be estimated 
  and visualized rapidly.} Shown are slices of the 3-D parameter space cube at fixed $q_{i}$ with the initial sampling of 21 states.
  Because the reference state is an uncharged particle, the error to charged parameters 
  is large. The lack of configuration space overlap causes the uncertainty in the error estimate 
  is large and unconverged. \subref{fig:n21-f6} and \subref{fig:n21-f31} show samples 
  of the free energy on either side of $q=0$. \subref{fig:n21-odd} shows a discontinuous 
  jump in uncertainty between the nearby charge $q=-1.52$ in \subref{fig:n21-f6}, 
  an artifact of poor configuration space overlap.
  Animated movies showing the full free energy 
  and uncertainty across the whole parameter space for initial and final samplings are included in the supplementary 
  materials.~\cite{NadenSupNote} Free energy is shown in units of $\kcalmol$.}
  \label{fig:3d-n21-slices}
\end{figure}

Even visualizing the thermodynamic properties and their uncertainties in 3-D space is a nontrivial task. 
Iterative determination of optimal states to sample is significantly harder.
 Fig.~\ref{fig:3d-n21-slices} 
shows the relative solvation free energy in the 3-D parameter space for three slices of fixed $q_{i}$, and 
samples drawn from the initial 21 states. Because the reference state is an uncharged particle, 
there is large uncertainty in the relative solvation free energy to charged particles, as seen by Fig.~\ref{fig:n21-f6} and 
Fig.~\ref{fig:n21-f31}. We show the entire 3-D space of solvation free energy as a 
function of the three force fields parameters in animated movie provided in the supplementary material.~\cite{NadenSupNote}

Regions of poor configuration space overlap and large uncertainty can be visually identified in the initial simulations. 
Fig.~\ref{fig:n21-f6} and Fig.~\ref{fig:n21-odd} are taken from two 
nearby values of charge. One would expect the uncertainty to change smoothly with the partial charges differing 
by only $0.08$, however the magnitude of the uncertainty changes by a factor of nearly two, 
causing a visual artifact. Unless there is some sort of phase change in the system (which, as it turns out, there is not), 
these sorts of artifact indicates that there is little configuration space overlap, and thus both the free energies themselves as well as the estimate of the uncertainties have not converged.
In order to improve our estimates, additional samples must be drawn 
at new parameter combinations to improve the configuration space overlap between our reference point and parameter regions such as the ones 
in Fig.~\ref{fig:n21-odd}. However, deciding exactly where to put these new samples cannot be easily done by 
visual inspection as in the 2-D case.  We must identify an algorithmic ways of placing these new points that can be easily automated, 
rather than having to do time-consuming manual trial-and-error.

\subsection{Adaptive sampling in 3-D parameter space and improving configuration space overlap}

As noted before, explicit thermodynamic paths need not be be directly defined by the user if property 
estimates are made by reweighting samples spanning the configuration space of all parameters of interest.
A multistate statistical analysis 
method such as MBAR~\cite{Shirts2008} takes into account the configuration space overlap from all 
samples relative to the state of interest. The concept of a single ``path'' is 
now obsolete since any two states are now connected through a network of 
configuration space overlap and connected states. 
Sufficient configuration space overlap between all of the states of interest 
and sampled states results in low statistical error in estimating relative properties 
between states on the network.

We define two types of configurational space overlap that help us better describe the issues of 
searching a multidimensional parameter space. 
We define the {\em local configuration space overlap} as the set of configurations shared between simulations performed at an arbitrary 
parameter combination, and other nearby parameter combinations up to some finite $\pm p$ 
away in parameter space, where $p$ is one of the parameters of interest. 
We also define {\em global configuration space 
overlap} as the extent to which all points of interest in the parameter space are connected 
to all other points in this parameter space through a connected network of regions 
with local configuration space overlap. 

A key issue in sampling multidimensional parameter spaces is that the most naive adaptive 
sampling will lead to local configuration space overlap, but not global configuration space 
overlap. The intuitive place to put additional samples to improve uncertainties in a free 
energy calculation might initially be the parameter combinations where uncertainties are the largest. 
Sampling only the largest uncertainty point may improve the local configuration space overlap
for states near this sampled state, but does not necessarily create a network of 
states with global configuration space overlap connecting the reference state and the state of interest.
Without a connected network, the uncertainty in relative solvation free energy differences 
between states inside the local neighborhood of states becomes smaller, but uncertainty 
to the reference state, or any other state outside of the local configuration space overlap, 
will still be large. The 
presence of local overlap but not global overlap is easy to identify when sampling 
along a 1-D path, since it is easy 
to tell where along the path there are insufficient samples.
Global overlap is harder to identify and create in higher dimensional space since it is much 
less obvious where new samples should be placed. We need not place samples in all places where they do not occur; 
we instead must identify approaches that can automatically construct a network of overlapping states.

In our previous studies,~\cite{Naden2014,Naden2014b} optimizing the
alchemical switches to identify the most statistically efficient pathways worked well to
optimize alchemical paths in one dimension.
Analyzing the sample variance in thermodynamic 
integration (TI) along paths and optimizing the $\hnln$ to reduced this variance identified the lowest error pathway. 
However, since we need a network of connected states in multidimensional space, optimizing TI would require 
reducing the multidimensional space to 1-D and connecting each state along a fixed path,
However, since there are many possible ways to connect states in multidimensional space, optimizing TI 
in this multidimensional space would require connecting each state to every other state along a fixed path, resulting in 
$(51^3)!$, more than $10^{10^{5}}$ possible overlapping paths, assuming the paths was restricted to visiting each state only once. 
We can instead create a collection of states that creates a global configuration space overlap network over the entire space
by sampling discrete states in the multidimensional space, then analyzing 
the configuration space overlap between pairs of states with MBAR.~\cite{Shirts2008}

We used an array of clustering and image recognition algorithms, outlined in further detail in the supporting information, to find clusters of local phase 
space overlap and connect them, gradually improving global configuration space overlap.  Local phase 
space overlap was identified in the $51^3$ grid by identifying adjacent points having nearly 
identical statistical uncertainty to the reference state. When leaving these regions, 
the estimate of the statistical uncertainty changes discontinuously. Lack of configuration space overlap between two clusters 
results in a nearly constant large uncertainty estimate between any two points in either cluster.
Treating the grid of uncertainty estimates to the reference state as a function of the three parameters, 
we interpreted this as a 3-D image.  We used  SciPy's~\cite{ScipyWeb} multi-dimensional image 
processing module, {\tt ndimage}, along with a density-based clustering algorithm, 
DBSCAN,~\cite{Ester1996} to identify clusters of grid points with local configuration space overlap based on 
estimated statistical uncertainty. 

We choose a new state to sample inside each cluster of local configuration space overlap.
Each cluster was treated as volume occupying shape in the parameter space, and 
each shape had a boundary identified as the parameter combinations on the 
border of the cluster. We chose one new state inside a given cluster to sample at a random location 
inside the cluster. We then drew a series of line connecting this new state to the reference 
state, and this new state to each other cluster's new state.
We identify the intersection of the series of lines with the cluster boundaries with SciPy's 
Sobel boundary detection algorithm.~\cite{Sobel1968} Sampling the additional point on the boundary of the 
cluster provides a way to extend the area of local configuration space overlap until 
clusters joined to create a network of configuration space overlap, eventually 
creating global configuration space overlap;  we detail 
how we choose point on the boundary below. 
We iteratively apply this until it creates
global configuration space overlap through a network of connected states. 

By using a variety of graph theory methods, we cam propose new states
at local configuration space boundaries that help to minimize the
number of sampled states required to get good sampling across the parameter space. 
We briefly summarize the algorithm here, with additional details of the algorithm covered in the
supplementary material~\cite{NadenSupNote} in 
% section~\ref{sec:AdaptiveDetails}. %xr
section~S.3. 
The
implementation used is available online.~\cite{NadenGithub}
We generated a complete, weighted graph
where the new states identified inside each cluster were the vertices, 
and each edge were the lines connecting the new states between clusters and 
the reference state.
The weight of each edge connecting the vertices was computed 
by numerical integration of the uncertainty at uniformly spaced points in Euclidean space on the edge.
The uncertainty of each integration point was estimated by multidimensional interpolation 
from nearby grid points since many integration points were not on the $51^3$ grid.
A minimum spanning tree (MST) was created from this complete weighted graph using Kruskal's 
algorithm~\cite{Kruskal1956} implemented in SciPy's sparse graph routines.
We sampled the vertices of the MST, residing inside a high uncertainty cluster, and the 
intersection of the MST's edges with the boundary of the cluster as the final set 
of states sampled in the next iteration of the algorithm.
This approach has the advantage of scaling to arbitrary $N$-dimensions, as direct
visualization of higher dimensional spaces becomes increasingly difficult. 

\begin{figure}[H]
  \centering
  \subfloat[]{\label{fig:n203-f38}\includegraphics{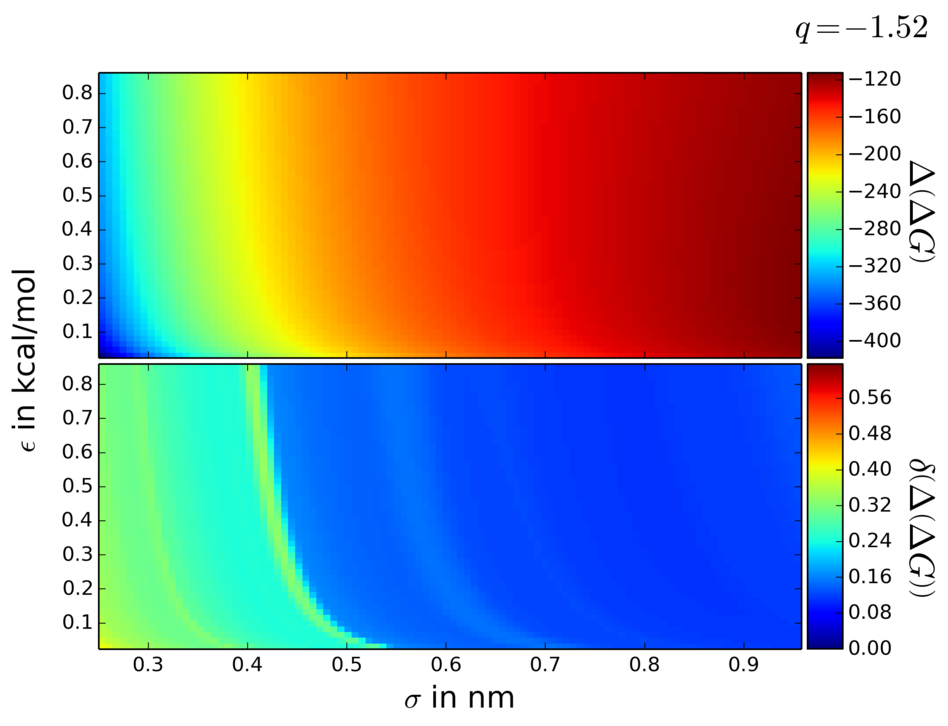} }~
  \subfloat[]{\label{fig:n203-odd}\includegraphics{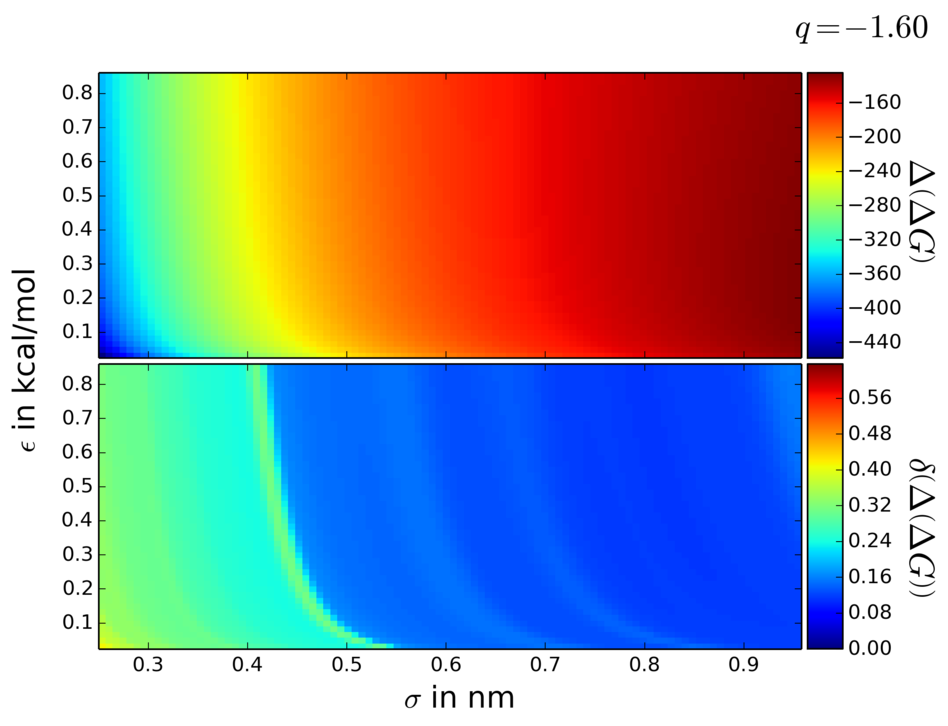}}\\
  \subfloat[]{\label{fig:n203-f21}\includegraphics{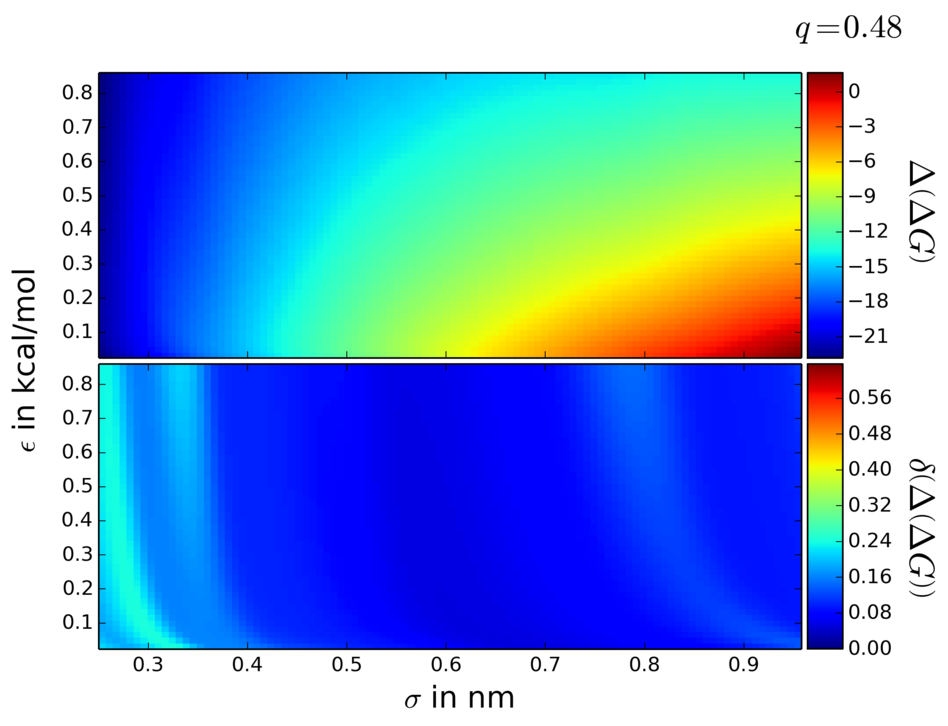}}\\
%   \subfloat[]{\label{fig:n203-f38}\includegraphics{images/smallSingleFrame_n203__f6.png} }~
%   \subfloat[]{\label{fig:n203-odd}\includegraphics{images/smallSingleFrame_n203__f5.png}}\\
%   \subfloat[]{\label{fig:n203-f21}\includegraphics{images/smallSingleFrame_n203__f31.png}}\\
  \caption[]{{\bf Adaptive sampling allows reduction of the uncertainty in the whole 
  multidimensional nonbonded parameter space.} Shown are slices of the 3-D parameter space 
  cube at the same fixed $q_{i}$ as in Fig.~\ref{fig:3d-n21-slices}. The total uncertainty 
  has been reduced by more than an order of magnitude with only a few adaptive iterations 
  and a total of 203 sampled states.
  \subref{fig:n203-f38} and \subref{fig:n203-f21} have significantly reduced error relative 
  to their counterparts in Fig.~\ref{fig:3d-n21-slices}. \subref{fig:n203-odd} no longer 
  has the discontinuous uncertainty as it did in Fig.~\ref{fig:3d-n21-slices}.
  Animated movies showing the full free energy 
  and uncertainty across the whole parameter space are included in the supplementary 
  materials.~\cite{NadenSupNote} Free energy is shown in units of $\kcalmol$.}
  \label{fig:3d-n203-slices}
\end{figure}

Statistical uncertainties in the free energy differences between states are reduced by two 
orders of magnitude even though the amount of sampling is only increased by nine times (21 to 204 states), 
because this adaptive algorithm generates good global configuration space overlap. Fig.~\ref{fig:3d-n203-slices} shows the same three 
slices of the 3-D parameter space as Fig.~\ref{fig:3d-n21-slices}, now with 203 sampled
parameter combinations, all adaptively chosen except for the initial 21 
combinations. We estimated properties at 101 $\sii$ points for the figure to improve image quality. All 
time comparisons are made assuming $51^3$ parameter combinations.
During this process, the maximum error in relative solvation free energy differences was reduced from  
$53.405\,\kcalmol$ to $0.631\,\kcalmol$ and the mean error was reduced from 
$16.162\,\kcalmol$ to $0.118\,\kcalmol$. However, the initial uncertainty is a misleading underestimate as 
much of the parameter space had no global configuration space overlap with the reference state, 
meaning the error estimates are unconverged.

\begin{figure}[H]
  \centering
  \includegraphics[width=0.45\textwidth]{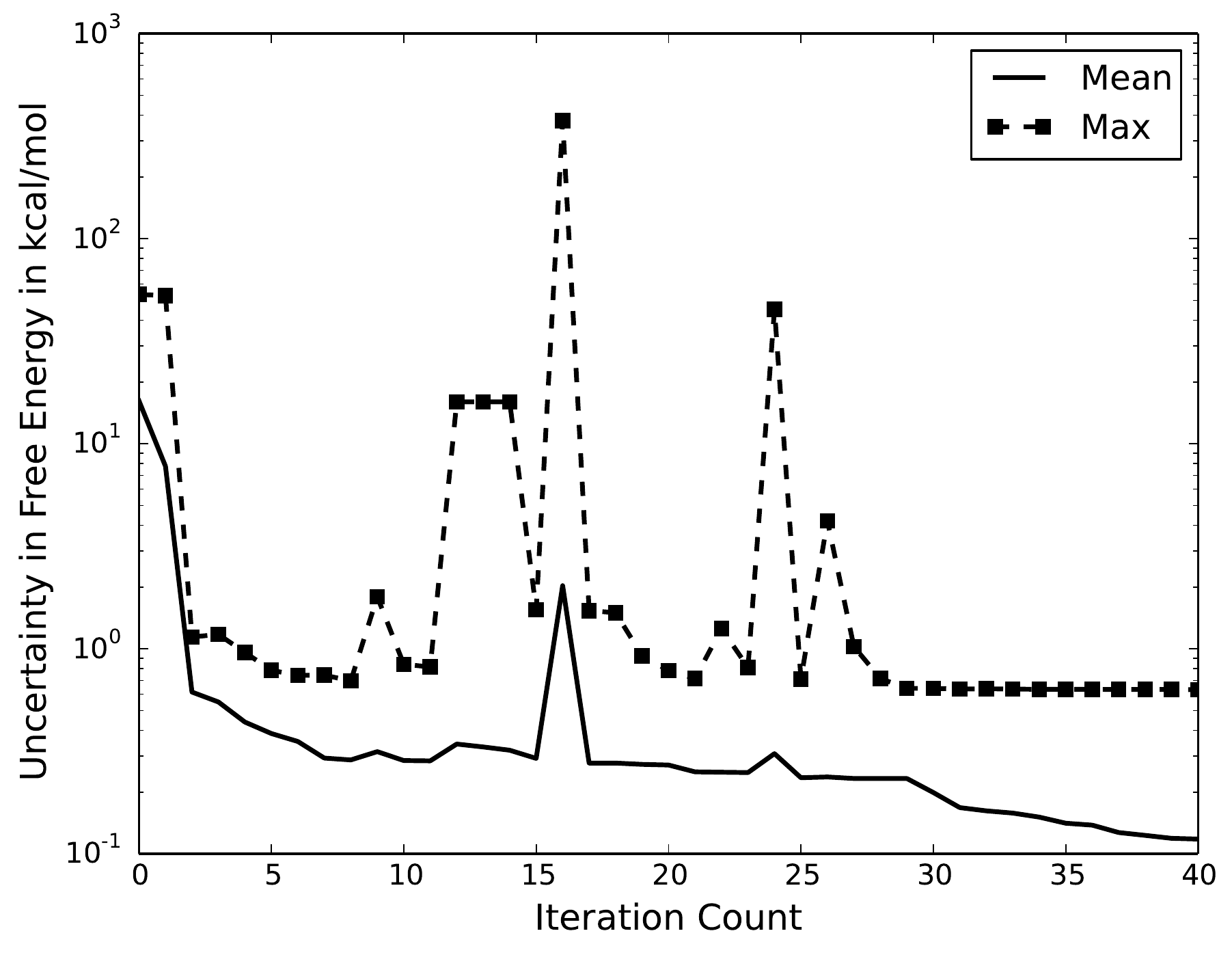}\\
  \caption[]{{\bf Discovery of regions of poor configuration space overlap appear as sudden jumps in the 
  maximum uncertainty from between two iterations, but these regions are joined by sampling adaptively.} The mean (solid) and 
  maximum (dashed) uncertainty in the free energy for the 3-D nonbonded parameter combinations 
  are plotted as function of iteration of the algorithm. The algorithm reduces the 
  uncertainty of the largest areas of uncertainty before moving to others, where it can 
  find regions of no configuration space overlap. Once some configuration space overlap is found through 
  adaptive sampling, the uncertainty jumps as a more converged estimate can be made. The 
  iterative process improves phase 
  space overlap and lowers overall uncertainty.
  Uncertainty in free energy is shown in units of $\kcalmol$ and shown using a logarithmic 
  scale to show changes at both large and small maximum uncertainty.}
  \label{fig:mean_max}
\end{figure}

The consequences of the poor configuration space overlap can be seen in 
Fig.~\ref{fig:mean_max} where the maximum and mean uncertainty jumps at a certain 
iterations. The jumps in uncertainty indicate that a new region of poor configuration space overlap 
has been identified and partially sampled. Ways to monitor when global configuration space 
overlap has been reached are discussed in Section~\ref{sec:converge}. 

\begin{figure}[H]
  \centering
  \includegraphics{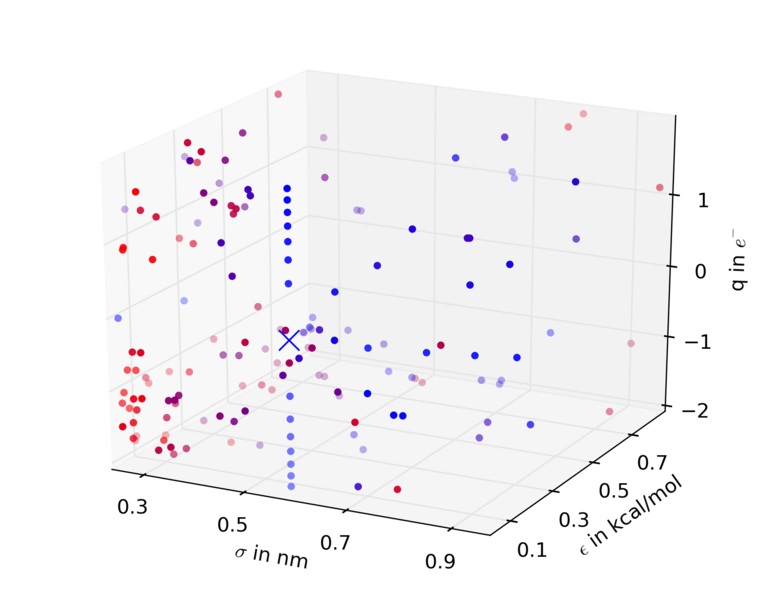}\\
  \caption[]{{\bf The adaptive sampling algorithm samples the entire parameter space.} 
  The sampled states are shown as a scatter plot with successive iterations moving from initial (blue)
  to final (red) as a function of the 3-D parameter space. Many new 
  states are selected at small $\sii$ as large uncertainty in the region is identified from 
  successive iterations. Fewer samples states are needed at intermediate $\sii$ as the 
  configuration space overlap to states already sampled is high. 
  The reference state is shown as an X.}
  \label{fig:scatter}
\end{figure}

The adaptive sampling algorithm correctly places samples to reduce regions of poor configuration space overlap. 
Fig.~\ref{fig:scatter} shows all of the 
sampled states in a scatter plot of the 3-D parameter space. 
Subsequent adaptive iterations are shown in color scale ranging from blue for the 
initial iterations to red in the final iteration. The large clustering of points is expected at small 
$\sii$, small $\eii$, and large $q_{i}$, because the water tightly rearranges around the 
particle due to very large Coulombic interactions. 
The tightly packed water arrangements share little 
no configuration space overlap with any other parameter combinations, so many samples at these 
states and connecting states are needed to accurately estimate properties. 

The bottleneck of computing the energies is completely removed with the linear basis function approach.
There are more than five times more states in the 3-D parameter (132,651) space than the 
2-D space (22,801). Despite this, computational cost to compute the energies in 3-D space only 
increases to less than 30 CPU seconds for 21 sampled states, and just over 50 CPU seconds 
for 203 sampled states. Taking again the conservative average of $1500$ CPU seconds 
needed per rerun, computing the energies at the 132,651 parameter combinations  
would have taken over 132 and 1280 CPU years for 21 and 203 sampled states, respectively.  
Optimized code to explicitly calculate the basis functions at the time of the 
force calculation would allow even faster vectorized calculations than the post-processing used here, 
allowing this method to 
scale to even larger multidimensional spaces. After removing this bottleneck, the main 
cost is in performing the 203 simulations at sampled states and running MBAR. Our simulations 
took an average of 25 CPU hours per simulation to run, and MBAR calculations took 108 CPU 
hours to compute properties. We would need to invest the time to run simulations 
and compute properties with MBAR, independent of how the energies were computed.  

The algorithm could be further optimized depending on the relative
cost of the simulations and of MBAR over very large numbers of
states. In this case, simulations were relatively cheap compared to
MBAR. If the simulations were more expensive, then shorter simulations
could be run between iterations. Additionally, more proposed states
for direct simulation could be generated at each step to reduce wall
time. For example, instead of 10 new states run for 10 ns, 20 new
states could be run for 5 ns each. In general, shorter cycles of
simulation plus analysis are expected to improve performance as within
well-sampled regions, the error in free energy estimates scales
approximately as $N^{-1/2}$. Adding significant configuration space overlap
between regions that are not connected will scale significantly
better.

\subsection{Computing Other Thermodynamic Properties and Comparing to Reported Results}

Estimating properties over the entire multidimensional parameter space at once 
can provide thermodynamic information which would otherwise require extensive simulations 
to compute at each thermodynamic state. This section looks at estimating five properties where
significant simulation is required when sampling each state individually: 
the relative solvation entropy and enthalpy, 
the absolute solvation free energy of ions, 
the radial distribution function (RDF) focusing on the first hydration shell, 
and the difference between the Born solvation free energy and the simulation estimate in 
the free energy of charging a particle.
In this section, we show how we can compute the enthalpy and entropy from the same 
collected data as used for free energies, compare ion parameter free 
energies to show accuracy and limitations in our approach, estimate RDFs without 
generating trajectories at the target parameters, and identify trends in the deviation 
of solvation free energies from the Born solvation approximation.

\subsubsection{Relative Solvation Entropy and Enthalpy}
We can estimate the relative solvation entropy and enthalpy alongside the relative solvation 
free energy without additional 
sampling. The relative solvation enthalpy is the difference in solvation enthalpy,~$\Delta H_{\text{solv}}$, 
between the reference state $X$ and any other state $j$ in the multidimensional parameter space. 
We compute the relative enthalpy difference, $\ddx{H_{Xj}}$, 
as
\begin{equation}
 \ddx{H_{Xj}} = \expect{\Delta U_{Xj}} \label{eq:enthalpy}\text{.} 
\end{equation}
Similarly the relative solvation entropy, $T\ddx{S_{Xj}}$, is computed as 
\begin{equation}
 -T\ddx{S_{Xj}} = \ddx{G_{Xj}} - \ddx{H_{Xj}} \label{eq:entropy}
\end{equation}
where $\ddx{G_{Xj}}$ 
is the relative free energy of solvation, and $\expect{\cdot}$ is the statistical expectation value, computed here by 
MBAR.~\cite{Shirts2008}
In all cases, we report the difference in thermodynamic properties with respect 
to the reference state. These properties we estimate have such large uncertainty from 
only the initial 21 states due to poor sampling that any number 
is essentially meaningless. Computing relative enthalpies and entropies generally 
requires significantly more samples to compute than relative free 
energies,~\cite{Paliwal2013b,Fenley2012,Fenley2015} as only samples with local 
phase space overlap to the end states contribute to precision of expectations of observables such as the enthalpy.

\begin{figure}[H]
  \centering
%   \subfloat[]{\label{fig:Hn203-f11} \includegraphics{images/smallSingleFrameH_n203_11.png} }~
%   \subfloat[]{\label{fig:Hn203-f24} \includegraphics{images/smallSingleFrameH_n203_24.png} }\\
%   \subfloat[]{\label{fig:Sn203-f11} \includegraphics{images/smallSingleFrameS_n203_11.png} }~
%   \subfloat[]{\label{fig:Sn203-f24} \includegraphics{images/smallSingleFrameS_n203_24.png} }\\
  \subfloat[]{\label{fig:Hn203-f11} \includegraphics{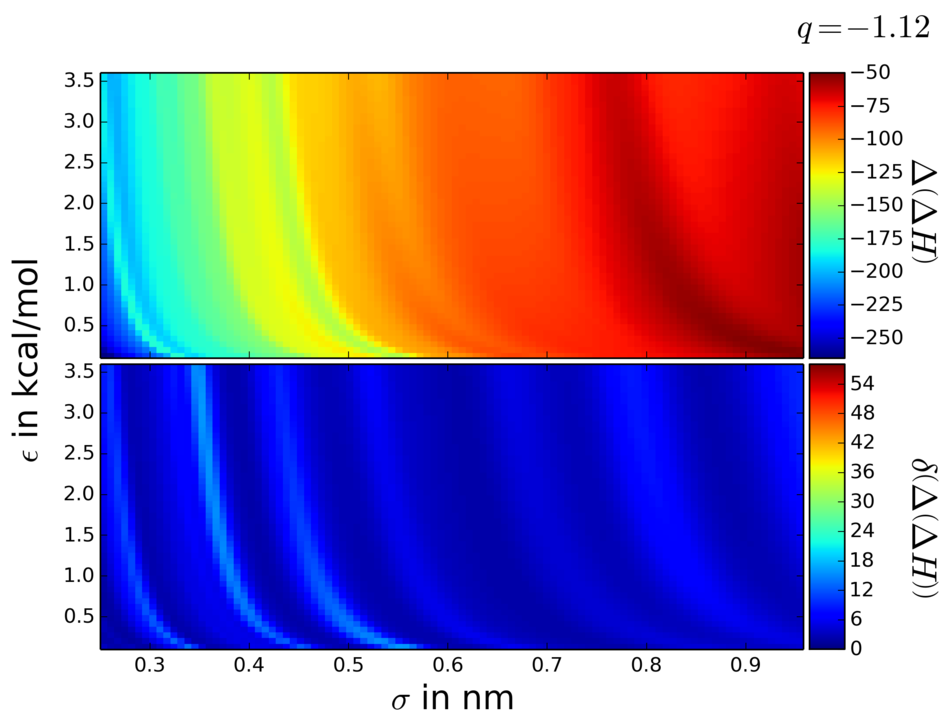} }~
  \subfloat[]{\label{fig:Hn203-f24} \includegraphics{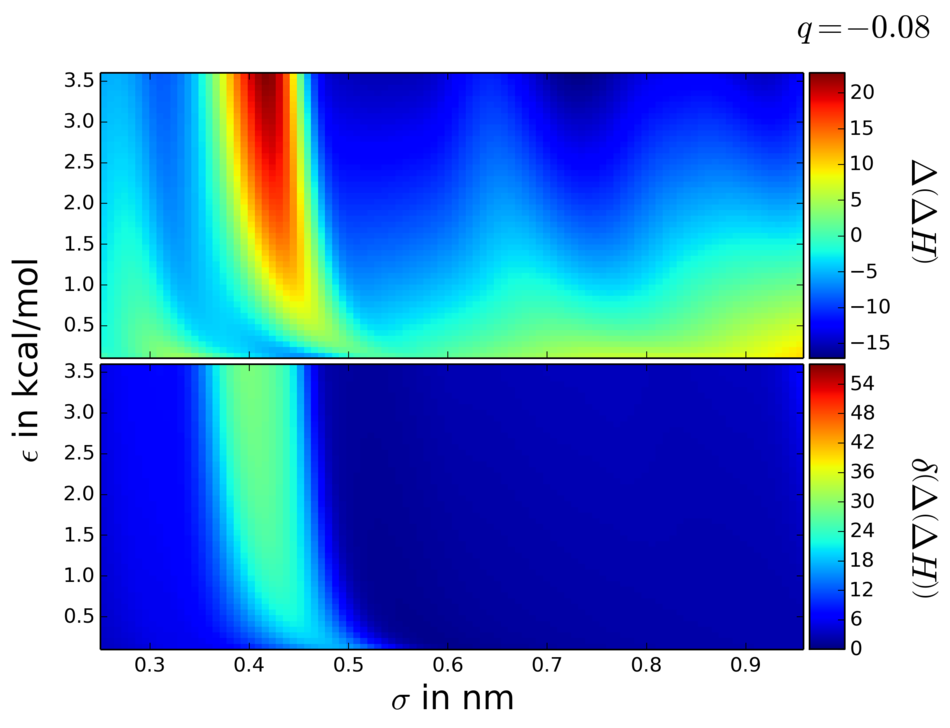} }\\
  \subfloat[]{\label{fig:Sn203-f11} \includegraphics{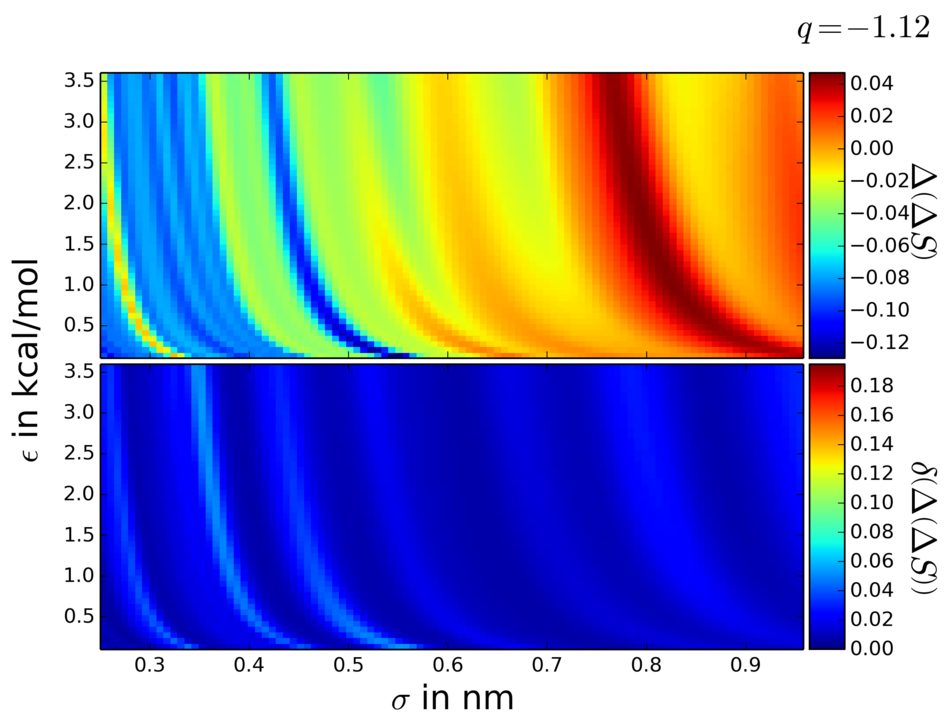} }~
  \subfloat[]{\label{fig:Sn203-f24} \includegraphics{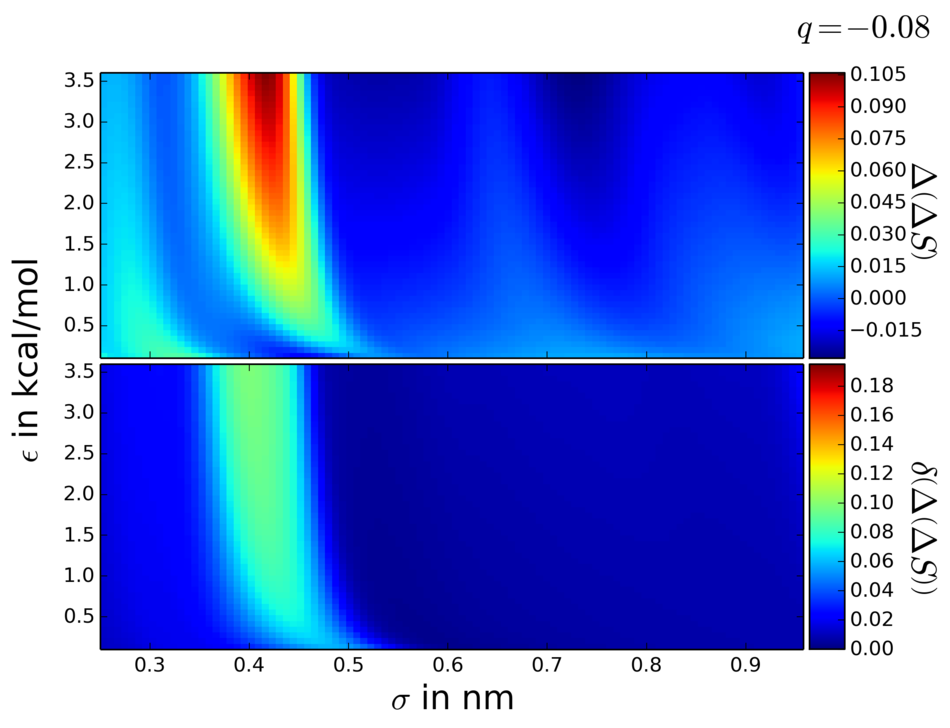} }\\
  \caption[]{{\bf Uncertainty in entropy and enthalpy are also reduced with the uncertainty 
  in the free energy.} The enthalpy $H$, (\subref{fig:Hn203-f11} and 
  \subref{fig:Hn203-f24}) and entropy $S$, (\subref{fig:Sn203-f11} and \subref{fig:Sn203-f24}) 
  are computed from 203 total sampled states and reported in kcal/mol. The uncertainty in both properties now 
  smoothly transitions between adjacent states. The uncertainty is still significantly larger than that of the 
  free energy, which is expected as these properties require more sampling to compute 
  accurately than the free energy. Additional samples or other means of computing these properties 
  would be required to reduce error further.} 
  \label{fig:HS-n203-slices}
\end{figure}

Additional sampling reduces the uncertainty in the estimates of relative solvation entropy 
and enthalpy by orders of magnitude, but not to the same extent as the uncertainty 
in the free energy. This is because whether or not there is good global phase space overlap does not ensure that 
a given state has good local phase space overlap with its neighbors.  
Fig.~\ref{fig:HS-n203-slices} shows the 
relative solvation entropy and enthalpy estimations, along with 
uncertainty at 203 sampled states. The uncertainty 
smoothly transitions between 
adjacent states, suggesting the estimates are numerically 
converged. However, the maximum uncertainty is several 
orders of magnitude larger than the relative solvation free energy. The maximum uncertainty
in relative solvation entropy changes from an uncertain estimate to $0.193\,\kcalmolK$, and the 
relative solvation enthalpy's maximum uncertainty drops to $57.5\,\kcalmol$. 
The mean uncertainty for relative solvation entropy and enthalpy fall to $0.0147\,\kcalmolK$ and $4.37\,\kcalmol$,
respectively. Although these error estimates are still too large to make practical predictions of solvation entropies, 
the estimates of the errors from the 203 states are well-defined, which is a marked improvement over the initial 
sampled states.  The whole configuration space has been sampled, and all states 
now have at least moderate local phase space overlap with its neighbors.  Once 
decent estimates of properties are found, we can run additional simulations 
on states that have the most desirable preliminary estimates of properties.

\subsubsection{Ion Solvation Free Energies \label{sec:ion_free_energy}}

We compare the results from our study to a detailed ion parameter study by Joung and Cheatham.~\cite{Joung2008} 
Their study parameterized Li$^{+}$, Na$^{+}$, K$^{+}$, Rb$^{+}$, Cs$^{+}$, 
F$^{-}$, Cl$^{-}$, Br$^{-}$, and I$^{-}$ in several water models including TIP3P 
and compared them to experimental and computational studies from others. They  
parameterized the ions based on several experimental observables, including 
free energy of solvation. We compared their absolute solvation free energies 
to absolute solvation free energies we computed from our method in Table~\ref{tbl:CompareResults}. 
The table shows the ion parameters, free energy of solvation, $\Delta G$, and 
first hydration shell (FHS) location estimated from their work and our approach. 
We compute absolute solvation free energies for our work by adding the free energies 
from the relative simulation to those of a single 
set of solvation simulations of the reference particle along a soft core 
potential~\cite{Beutler1994,Zacharias1994} and a 1-1-6 
parameterization.~\cite{Pham2011,Naden2014} These simulations were run with the 
same conditions as described in section~\ref{sec:exp_design} at 11 states uniformly 
distributed along $\plam$ from 0 to 1 of the soft core path. 

We re-computed parameter values and adjusted the reported free energies to 
make the values from Joung and Cheatham~\cite{Joung2008} comparable to this work.
Their ion $\sii$ was calculated for Lorentz-Berthelot mixing rules. We back calculated 
the $\sii$ for geometric mixing rules in Table~\ref{tbl:CompareResults} by setting the $\sij$ between 
the ion and the oxygen in water equal in both mixing rules, giving the relation
\begin{equation}
 \sigma_{\text{ion-ion, geo}} = \frac{\left(\sigma_{\text{ion-ion, LB}} + \sigma_{\text{OW-OW}}\right)^2}{4\sigma_{\text{OW-OW}}}
\end{equation}
where $\sigma_{\text{ion-ion, geo}}$ is the reported ion $\sii$ in Table~\ref{tbl:CompareResults},
$\sigma_{\text{ion-ion, LB}}$ is $\sii$ for the Lorentz-Berthelot mixing rule reported by 
Joung and Cheatham, and $\sigma_{\text{OW-OW}}$ is the TIP3P oxygen-oxygen $\sjj$ 
which we assumed was constant between the mixing rules. The solvation free 
energies from Joung and Cheatham have been adjusted by -1.9 $\kcalmol$ to 
remove the correction they added for ideal gas expansion when comparing simulations, 
carried out with gas-phase standard states at 1 M, to experimental results, 
where gas-phase standard states are typically 1 atm.~\cite{Joung2008,Grossfield2003}
\begin{landscape}
\begin{table}[H] %Free energies, direct and indirect
 \caption{
  Absolute solvation free energies and first hydration shell (FHS)
  locations compared from this work and from Joung and Cheatham.~\cite{Joung2008}
  Ion $\sij$ were back calculated from Lorentz-Berthelot to geometric mixing 
  rules for ion/oxygen interactions. $\eii$ and $\Delta G$ are in $\kcalmol$; 
  $\sii$ and FHS location are in nm. Error in this work's FHS computed by 200 bootstrap samples~\cite{Efron1993}
  with a discretization of $\pm 0.0075$ nm.
\label{tbl:CompareResults}}
\begin{center}
\begin{tabular}{|c|c|c|c|c|c|c|}
\hline
\multirow{2}{*}{Ion}	& \multirow{2}{*}{$\sii$}	& \multirow{2}{*}{$\eii$}	& \multicolumn{2}{|c|}{$\Delta G$}		& \multicolumn{2}{|c|}{FHS} 		    \\
			&				&				& Joung and Cheatham	& This Work		& Joung and Cheatham	& This Work	    \\
\hline
Li+			& 0.1965			& 0.0290			& -115.6		& -105.26 $\pm$ 0.54	& 0.196			& 0.211	$\pm$ 0.016 \\
Na+			& 0.2479			& 0.0874			&  -90.6		&  -90.63 $\pm$ 0.44	& 0.238			& 0.234	$\pm$ 0.015 \\
K+			& 0.3039			& 0.1937			&  -72.6		&  -72.42 $\pm$ 0.36	& 0.275			& 0.279	$\pm$ 0.016 \\
Rb+			& 0.3231			& 0.3278			&  -67.6		&  -67.31 $\pm$ 0.35	& 0.292			& 0.294	$\pm$ 0.030 \\
Cs+			& 0.3532			& 0.4065			&  -62.5		&  -62.13 $\pm$ 0.35	& 0.311			& 0.309	$\pm$ 0.021 \\
F-			& 0.4176			& 0.003364			& -121.6		& -120.91 $\pm$ 0.45	& 0.263			& 0.257	$\pm$ 0.016 \\
Cl-			& 0.4617			& 0.0356			&  -91.5		&  -90.99 $\pm$ 0.36	& 0.313			& 0.309	$\pm$ 0.028 \\
Br-			& 0.4825			& 0.0587			&  -84.8		&  -84.48 $\pm$ 0.36	& 0.329			& 0.333	$\pm$ 0.015 \\
I-			& 0.5396			& 0.0537			&  -75.9		&  -75.89 $\pm$ 0.34	& 0.351			& 0.347	$\pm$ 0.017 \\
\hline
\end{tabular}
\end{center}
\end{table}
\end{landscape}

The solvation free energies from our work and Joung and Cheatham's work are within statistical 
error. 
The free energies from this study appearing in Table~\ref{tbl:CompareResults} are within 
two standard deviations of Joung and Cheatham~\cite{Joung2008} for all 
ions except Li$^{+}$, which is outside the parameter range studied. 
The comparable accuracy of our results to those of 
Joung and Cheatham provide validation for the free energies we report. Our method 
has the added benefit of computing the solvation free energy for arbitrary 
parameter combinations. The ability to compute properties for arbitrary parameter 
combinations comes from sampling only 203 states plus 11 for the absolute solvation 
free energies. Joung and Cheatham carried out 12-13 simulations for each of the 
9 ions, resulting in 108-117 simulations for comparison. We were able to compute 
properties at roughly 14,000 times the number of parameter combinations, for only double the 
simulation cost.

Our method breaks down if the parameter combination falls outside the defined 
range. The parameters 
for the Li$^{+}$ ion in Table~\ref{tbl:CompareResults}
fall significantly outside the range we searched, namely, the $\sii$ is less than the 0.25 nm 
minimum. The estimated free energy for this parameter combination is not within 
statistical error for that of Joung and Cheatham.~\cite{Joung2008} The estimated 
value for any thermodynamic property at this ion will likely inaccurate as the 
estimation is now an extrapolation instead of a thermodynamically-consistent interpolation between sampled 
states. Estimates on parameter combinations falling just outside the searched range, such 
as the Na$^{+}$ ion, appear to still be accurate, so the range of convergence of these calculations outside of samples parameters is not zero.

\subsubsection{Estimating Radial Distribution Functions}

We can estimate the radial distribution function (RDF or $g(r)$) of 
a specified parameter combination without explicitly sampling that combination. 
The first hydration shell and the water 
RDF are properties that many have tried to compute 
accurately and compare to experiment.~\cite{Joung2008,Jensen2006,Aqvist1990,Beglov1994,Smith1994,Dang1992,Dang1994,Dang1995,Dang1993}
Traditionally, a RDF is generated by measuring the distances between two specified 
atomic groups (e.g. ion-water, water oxygen-water oxygen, etc.) 
generated over a trajectory, counting the number of pairs that are within a shell 
of size $r+\delta r$, then average over the shell volume and whole trajectory. The fact that the 
RDF is an average property and dependent only on the configuration
implies it is a thermodynamic equilibrium property 
which can be computed as a statistical observable. 
Computing an equilibrium expectation value of some observable, A, 
with MBAR~\cite{Shirts2008} is 
\begin{equation}
 \expect{A}_a = \sum_{n=1}^{N} W_{na}A(\bx_{n})
\end{equation}
where the summation runs over all samples in all states, $W_{na}$ are the 
statistical weights from reweighting each $n$th drawn sample in the $a$th state, 
the $A(\bx_{n})$ are the observed values from sample $n$ and are functions of the configuration, 
$\bx$. The weight MBAR assigns to each sample $n$ is~\cite{Shirts2008}:
\begin{equation}
W_{na} = \frac{e^{f_a - u_a(\bx_{n})}}{\sum_{k=1}^K N_k e^{f_k-u_k(\bx_n)}}
\label{eq:weight}
\end{equation}
where $f_a$ is the reduced free energy ($\beta A$ or $\beta G$) of
state $a$, $u_a(\bx_n)$ is the reduced internal energy ($\beta U$) of
the configuration $\bx_n$ in state $a$, and $N_k$ is the number of
samples collected from state $k$.
The observable for computing the RDF is the discrete count of pairs within a specified 
$r+\delta r$ shell normalized by the shell radius and the number volume 
$\rho = N_{\text{particles}}/V$. We must estimate the RDF at multiple shell volumes in 
order to generate a complete RDF curve.

\begin{figure}[H]
  \centering
%   \subfloat[Cl$^{-}$]{\label{fig:rdf-Cl}\includegraphics[width=0.45\textwidth]{images/rdfCl-.eps}}~
%   \subfloat[Li$^{+}$]{\label{fig:rdf-Li}\includegraphics[width=0.45\textwidth]{images/rdfLi+.eps} }\\
%   \subfloat[Cl$^{-}$]{\label{fig:rdf-Cl}\includegraphics[width=0.45\textwidth]{images/figure8a.eps}}~
%   \subfloat[Li$^{+}$]{\label{fig:rdf-Li}\includegraphics[width=0.45\textwidth]{images/figure8b.eps} }\\
  \subfloat[Cl$^{-}$]{\label{fig:rdf-Cl}\includegraphics[width=0.45\textwidth]{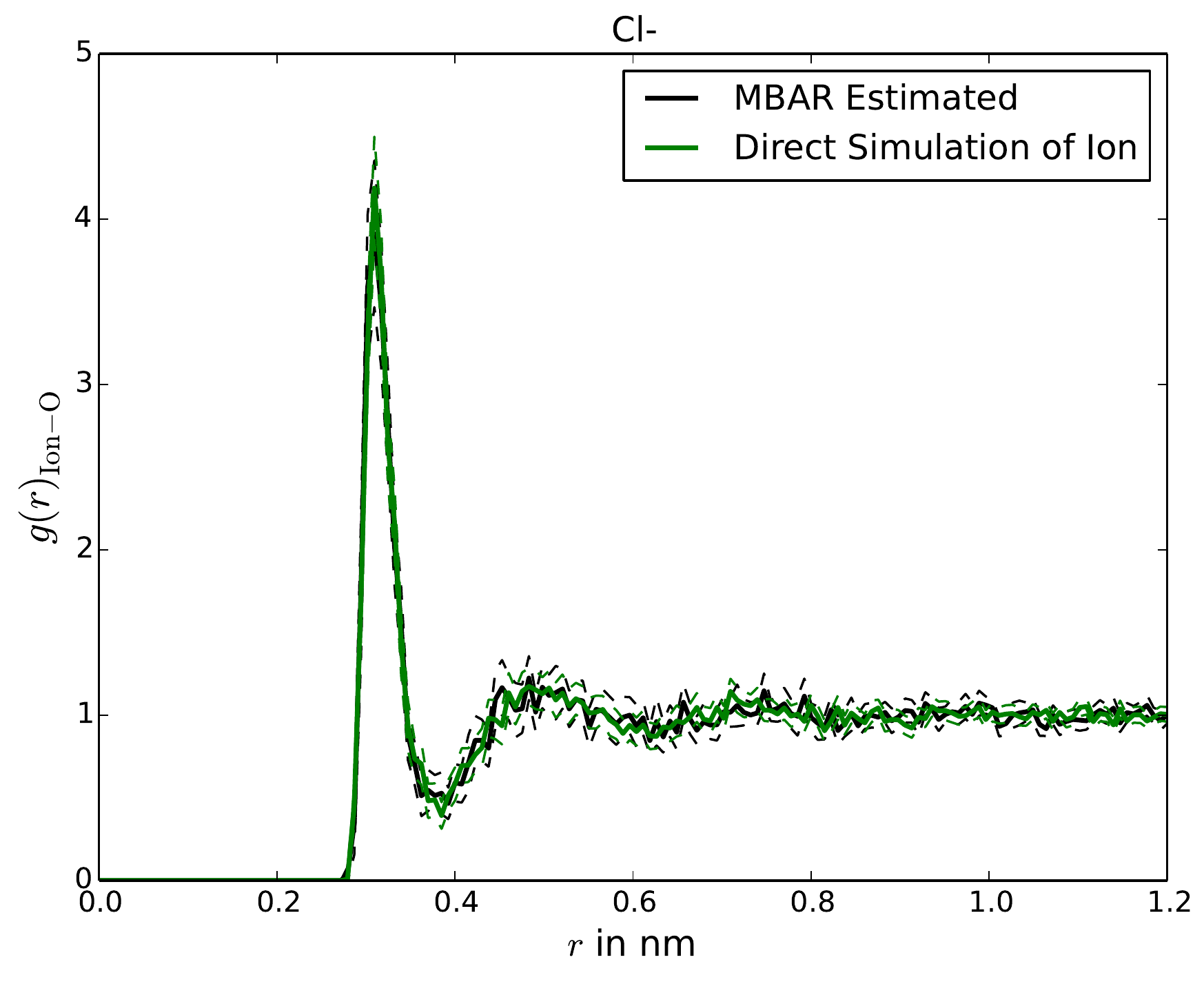}}~
  \subfloat[Li$^{+}$]{\label{fig:rdf-Li}\includegraphics[width=0.45\textwidth]{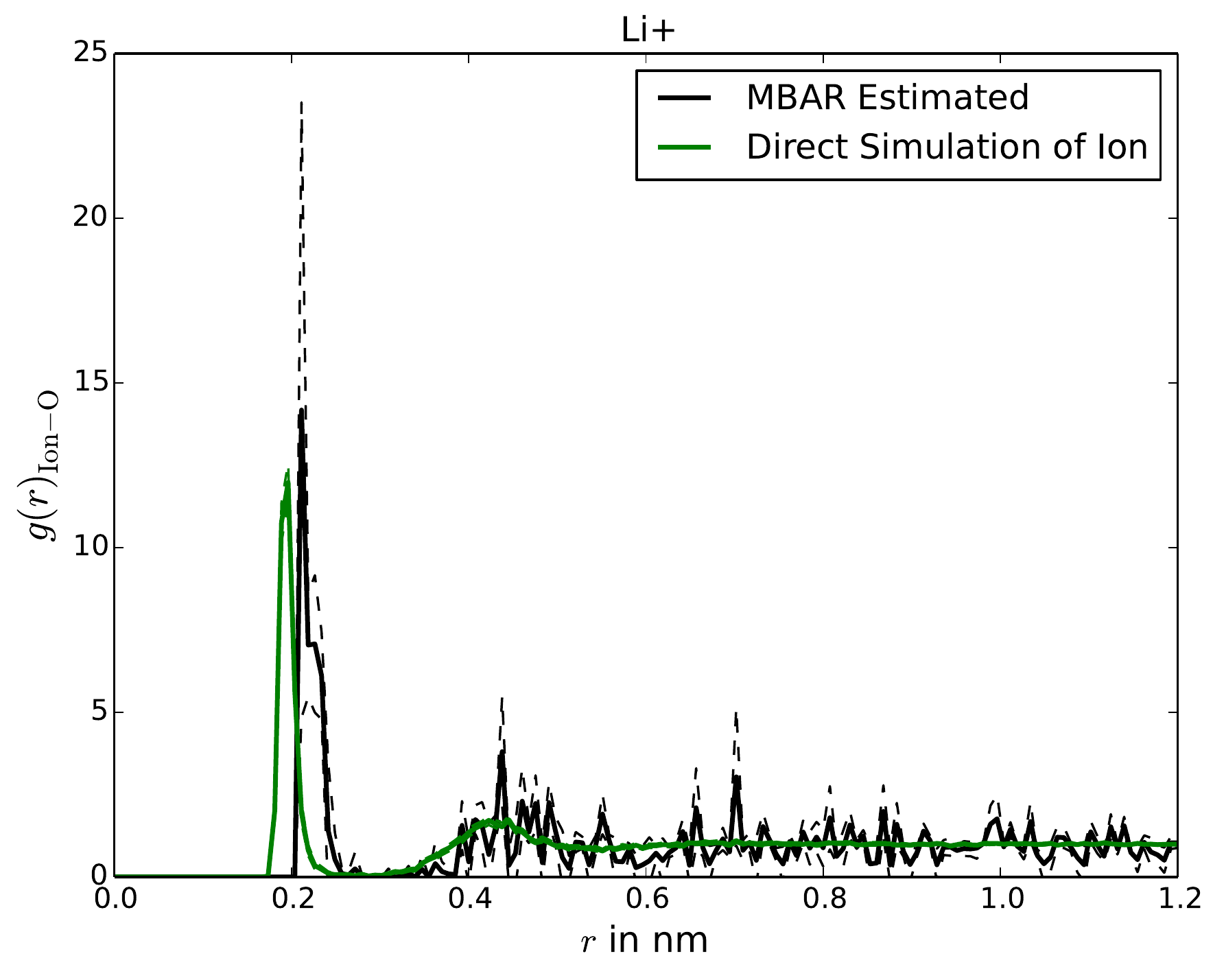} }\\
  \caption[]{{\bf Radial distribution functions (RDF) can be estimated at any 
  parameter combination inside the parameter search range.}  
The ion-oxygen RDF in TIP3P water is shown for
  Cl$^{-}$ and Li$^{+}$ with Joung and Cheatham parameters.~\cite{Joung2008}
  The RDFs are estimated without sampling the explicit parameter combinations
  using 160 discrete bins.  Estimates 
  made from parameters inside the parameter range, Cl$^{-}$ \subref{fig:rdf-Cl} 
  are accurate as the first hydration shell is predicted within error 
  to Joung and Cheatham.~\cite{Joung2008} in Table~\ref{tbl:CompareResults}. 
  Estimating parameters which fall outside the searched 
  parameter range, Li$^{+}$ \subref{fig:rdf-Li}, are inaccurate.
  Error is shown dashed lines of two standard deviations, computed by MBAR for the black 
  curves and 200 bootstrap samples for the green curves.}
  \label{fig:rdf}
\end{figure}

The first hydration shells (FHS) are accurately predicted by the RDF estimation. Fig.~\ref{fig:rdf} 
shows the RDF estimated for the Li$^{+}$ and the Cl$^{-}$ ions from 
Joung and Cheatham.~\cite{Joung2008} The RDF is estimated at 160 discrete bins (0.0075 nm spacing) 
from $r=0$ to $r=1.2$ nm along with the error in that estimate. The black curves  in Fig.~\ref{fig:rdf} are the 
estimates from MBAR computed purely from data collected at our 203 states, and not from 
any data drawn at the ion's parameters. Error in the MBAR estimate is shown as 
dashed lines and is two standard deviations of the uncertainty in the RDF, also 
computed by MBAR. To validate the MBAR results, the green curves in Fig.~\ref{fig:rdf} show the 
RDF computed from simulations at the given ion's parameters. The data from these direct simulations of the ions 
were not used in the MBAR estimate. The error in the green curves is taken from 
200 bootstrap samples~\cite{Efron1993} of the RDF for each ion. The Li$^{+}$ ion RDF is shown in Fig.~\ref{fig:rdf-Li} to again emphasize that 
estimates made outside the parameter range tend to break down, evidenced by the 
erratic behavior in the RDF. 

The Cl$^{-}$ ion in Fig.~\ref{fig:rdf-Cl} shows an example where we can estimate solvation structure 
and improve future simulations. 
We further validated the RDF calculation by determining the peak of the first hydration 
shell for every ion from Joung and Cheatham as the bin with highest occupancy~\cite{Joung2008} and compared our 
results to theirs in Table~\ref{tbl:CompareResults}. The peaks we estimate and 
those from Joung and Cheatham are in agreement with each other, within 
the error of our bin width of 0.0075 nm.  
The RDF curves generated using reweighting are not as smooth as the RDF curves from other 
studies, and we do not expect them to be smooth with the range of parameters we searched. 
However, all features are well preserved. This approach can be used to broadly search parameter 
space and generate approximate RDF's, until those that 
replicates the RDF or hydration shell properties of interest are identified. Further simulations 
could then be run around the sets of properties which gave the RDF replicating the target 
properties to make more accurate estimates, resulting in searches over a much narrower 
parameter space than examined through here. A complete set of the RDFs estimated via reweighting and direct simulation 
for every ion is included in the supplementary material.~\cite{NadenSupNote}

\subsubsection{Born Approximation to Solvation Free Energy \label{sec:Born}}
The Born approximation to solvation free energy measures the effort to transfer a charged particle 
between two dielectrics. The free energy differences for this approximation of 
transferring a hard sphere particle between vacuum and a fluid is
\begin{equation}\label{eq:Born}
 \Delta G = \frac{q^{2}}{4\pi\epsilon_{0}R_{ij}}\left(\frac{1}{\epsilon_{d}} - 1\right)
\end{equation}
where $\epsilon_{d}=92$ is the estimated dielectric constant of our fluid, TIP3P water~\cite{Lamoureux2003},  
and $R_{ij}$ is the Born radius. 

We can estimate the Born radius of any particle in our search space from our 
sampled states. Choosing the correct Born radius, or effective 
hard sphere (EHS) radius is a nontrivial task. However, we can estimate the 
EHS radius with our RDF calculation. We first compute the RDF for a given 
parameter combination, $g(r)$, then compute the EHS radius by determining an
$r_{0}$ where the following conditions for the oxygen-ion g(r) are met:
\begin{alignat}{2}
 g(r_{0} - \delta r) &= 0 \\
 g(r_{0})            &= 0 \\
 g(r_{0} + \delta r) &> 0
\end{alignat}
to a tolerance of $10^{-5}$. $r_{0}$ can be interpreted as the point on $g(r)$ 
where the probability of finding a particle changes from zero to nonzero. 
We set $R_{ij} = r_{0}$ for the Born solvation calculations.

We applied correction terms to our estimated free energies to remove errors 
introduced by our choice of simulation settings and water model. These corrections 
allow a comparison of free energy between different methods without having 
a methodological dependence. All of the corrections we applied are detailed 
in H{\"u}nenberger and Reif~\cite{Hunenberger2011} and we go through explicit 
detail of which corrections we applied and why in the supplementary 
material~\cite{NadenSupNote} in 
% section~\ref{sec:g_correction}. %xr
section~S.6.

There are a number of reasons the the Born approximation is not
perfect for our particle-water system. These imperfections come from
not simulating a truly infinite medium, the water model having an
asymmetric charge distribution, Lennard-Jones terms affecting the
free energy of transfer, and the fact that $R_{ij}$ may change on
charging since we have soft particles. We are
interested in identifying deviations from the Born approximation with
our method, given that we fully expect deviations from these
imperfections.

The Born approximation to the solvation free energy is the free energy of transferring a charged
hard sphere with radius $R_{ij}$, not for the free energy of solvating
the uncharged particle. To remove this dependence on the cavitation free energy,~\cite{Hunenberger2011}
we we estimate the $R_{ij}$ from the RDF as described
above for each uncharged combination of $\sij$ and
$\eij$, then calculate the free energy difference to the same
values of $\sij$ and $\eij$ but with a charge. This allows us to compare 
our free energy of charging to the Born approximation to the free energy of 
charging, and identify deviations between the model and our simulation. We do
not recalculate $R_{ij}$ at the end state.

\begin{figure}[H]
  \centering
%   \subfloat[]{\label{fig:B203-f10}\includegraphics{images/smallBornSingleFrame_n203__f10.png} }~
%   \subfloat[]{\label{fig:B203-f40}\includegraphics{images/smallBornSingleFrame_n203__f40.png}}\\
  \subfloat[]{\label{fig:B203-f10}\includegraphics{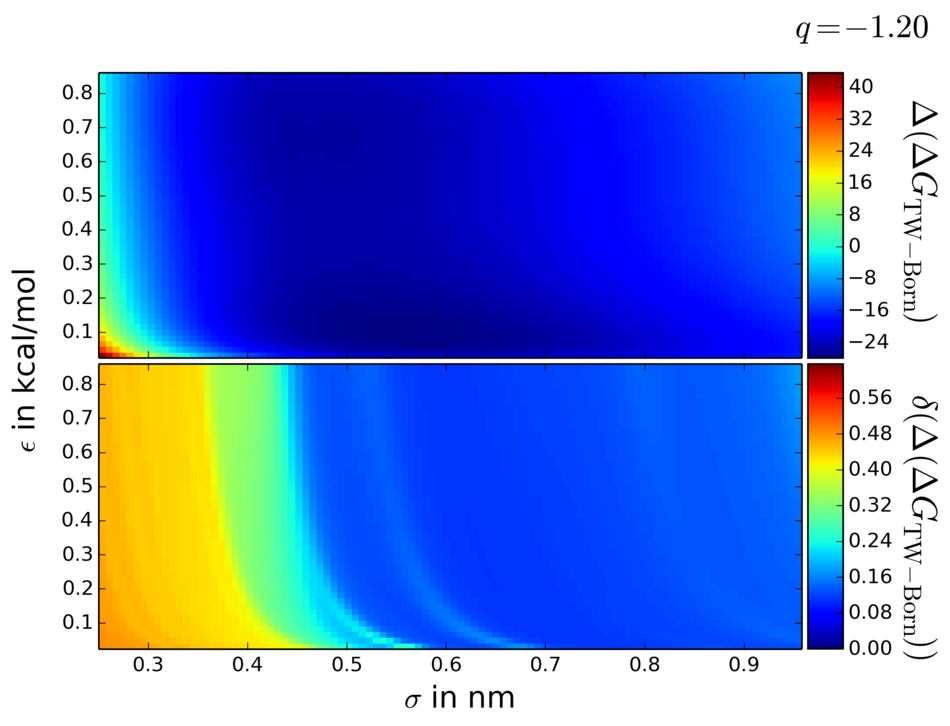} }~
  \subfloat[]{\label{fig:B203-f40}\includegraphics{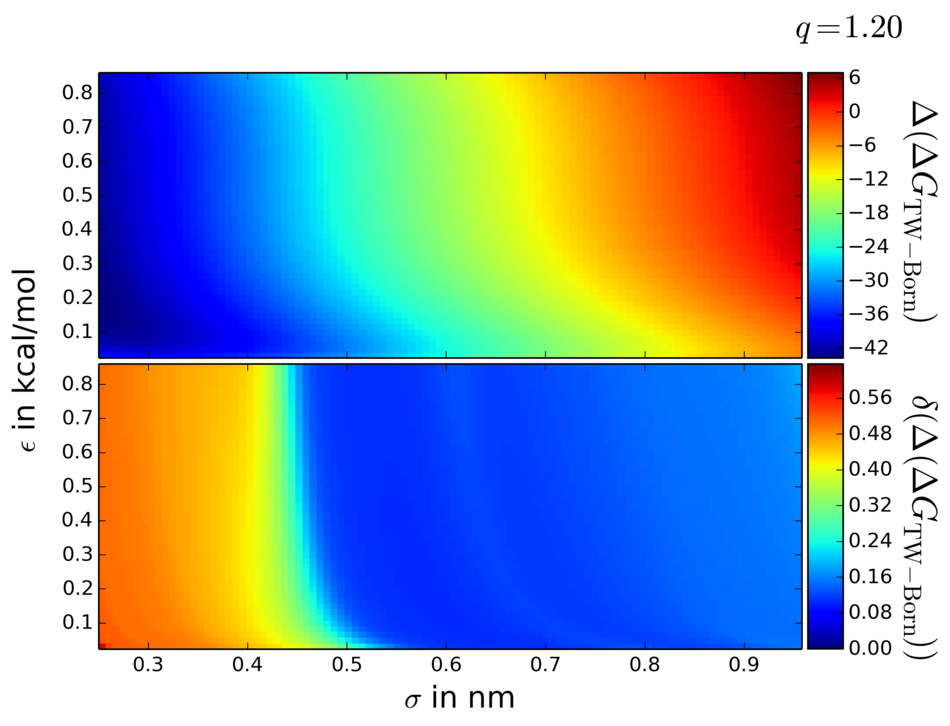}}\\
  \caption[]{{\bf Trends and failures in approximations can be visualized over 
  wide parameter space.} 
  The deviation of the Born hydration free energy from the computed solvation free energy is shown for two fixed 
  slices of $q_{i}$. Explicitly shown is the Born free energy minus this work's free energy 
  estimates: $\Delta G_{\text{Born}} - \Delta G_{\text{TW}}$. Free energy difference 
  estimates for each combination of $\sii$ and $\eij$ are relative to the 
  same combination at $q_{i}=0$ as to approximate only the contribution of 
  charging a given sphere in solvent. \subref{fig:B203-f10} and \subref{fig:B203-f40} 
  show the deviations with the solute carrying a $\pm q$ charge. The Born model generally 
  predicts a more favorable interaction relative to the simulation. An exception to this trend is at very small $\sii$ and $\eii$ for a 
  negatively charged particle where it predicts a less favorable interaction than the simulation 
  as the TIP3P water hydrogens can tightly pack around the 
  particle.
  Animated movies showing the full free energy 
  and uncertainty across the whole parameter space are included in the supplementary 
  materials.~\cite{NadenSupNote} Free energy is shown in units of $\kcalmol$.}
  \label{fig:3d-B203}
\end{figure}

Both trends and failures of the Born approximation can be easily visualized for the entire 
parameter space. Fig.~\ref{fig:3d-B203} shows the difference between the Born 
approximation ($\Delta G_{\text{Born}}$) and this work's estimate for the charging  
free energy ($\Delta G_{\text{TW}}$). Any deviation from 
$\Delta\Delta G_{\text{TW-Born}} = 0$ indicates nonidealities relative to Born approximation 
to the free energy of charging. There are several 
deviations which can be seen in the figure. The first deviation is the Born free energy 
generally predicts less favorable solvation free energy for both signs on the charged particles 
($\Delta G_{\text{TW}} < \Delta G_{\text{Born}} < 0 $). However, this deviation 
is asymmetric as the deviation in from Born theory of the positively charged particle at 
$q = +2$ is up to $-152.1\,\kcalmol$, but the negatively charged particle only deviates 
up to $-78.5\,\kcalmol$ at $q=-2$. 
The charging free energy 
more strongly depends on $\sii$ for the positive particle as the $\ddx{G}$ in 
Fig.~\ref{fig:B203-f10} spans 13.7 $\kcalmol$
from $\sii \approx 0.5$ to $\sii \approx 0.95$ on average, whereas in Fig.~\ref{fig:B203-f40} 
the span 27.6 $\kcalmol$ on average
in the same range of $\sii$. We can also observe the opposite case where the 
Born estimate is more favorable than the observed estimate ($\Delta G_{\text{Born}} < \Delta G_{\text{TW}} < 0 $) 
in Fig.~\ref{fig:B203-f10} where $\ddx{G} > 0$ at small $\sii$ and $\eii$. 
This opposite case occurs because the negatively charged particle attracts 
the water's hydrogens, which do not have Lennard-Jones interactions, and would 
be able to approach much more closely than the Born model predicts with its hard 
sphere approximation. Simply estimating free energy at a few states in the parameter space would have 
been insufficient to observe these broad trends as a significant degree of interpolation, 
or worse extrapolation, would have been required.

A full animation showing $\Delta\Delta G_{\text{TW-Born}}$ at combination in 
the parameter space is included in the supplementary material.~\cite{NadenSupNote}

\subsection{Monitoring numerical bias \label{sec:numerics}}

The numerical bias caused by the process of calculating energies from perturbations of reference states 
can be minimized.
Eq.~\eqref{eq:xform-total-delta} and in the supplementary material
% Eq.~\eqref{eq:h_Zi} %xr
Eq.~S.5
involve the addition and subtraction of many small numbers, depending 
on reference states $X$ and $Y$. This can lead to rounding errors which may propagate to the 
simulation package and be made worse by the software's precision. Choosing reference 
states with larger $\Delta C_{n,XY}$ can help 
reduce accumulation of error. If the software natively allows access to the basis function 
values, then this source of numerical error is eliminated. However, we must quantify our 
numerical error here since the perturbation approach was chosen.

We find that rounding errors do not propagate from the perturbed basis function representation to the 
thermodynamic properties for these calculations. The rounding 
errors for this system were checked by evaluating the energy of each sampled configuration
at every sampled state as though we did not use the basis function approach.
The energies of every configuration evaluated at every sampled state computed from directly from GROMACS
were compared to the energies computed 
from the basis functions and any deviation was a result of numerical bias. 
The energies computed from basis function calculations and the energies directly from GROMACS reruns differed by less than $0.002\%$.
This very small relative error does not assure that errors themselves are negligible, since large energies have large 
absolute errors. The largest absolute
error in all 203 simulations was was $11.804\,\kcalmol$, which at first appears is likely to have a significant effect in the final answers.
However, these large absolute errors do not affect any of the property estimates. This is because these large rounding 
errors occur when the trajectory from a particle with small $\sii$ is evaluated 
in a force field for a particle with large $\sii$. This often resulted in the oxygen of TIP3P water 
 being within the large particle's excluded volume, resulting in a highly repulsive interaction.
Every configuration with rounding error in this study had 
a Boltzmann weight, $\exp\left(-\beta U(\vecr,\plam)\right)$, indistinguishable from zero at 
machine precision, and thus these errors do not contribute to any of the properties of interest.
The energies calculated using reference states thus give results that are sufficiently 
close to those from direct evaluation of the energies for all uses.

\subsection{Convergence and alternate algorithm conditions\label{sec:converge}}

Examining the uncertainty estimate from the reference state alone is insufficient to determine 
convergence of the calculations. The multidimensional space initially has many regions of
configuration space 
overlap, which causes unconverged estimates of the properties and their uncertainties. 
Poor overlap implies that the mean 
and the maximum uncertainty alone are not appropriate gauges for convergence since 
the error does not consistently decrease with number of samples as observed in Fig.~\ref{fig:mean_max}.
As discussed above, a network of overlapping configuration space between all states required for accurate 
estimate of properties and quantification of the uncertainties in these properties, and we need a way to diagnose whether this 
network has been created at a given stage of the simulation.

The configuration space overlap can be analyzed through a multidimensional extension of the Overlapping 
Distribution Method.~\cite{Frenkel2002,Pohorille2010} This method can quantify the 
overlap between states by considering the probability of each sample occurring in every state. 
The unnormalized probability of a sample can be computed from 
its Boltzmann weight. 
Just as each sampled configuration carries a Boltzmann weight 
for the state in which it was drawn, the configurations can be reweighted to all other 
states to determine what the relative Boltzmann weights are in the other sampled states.~\cite{Bennett1976,Kumar1992,Kumar1995,Shirts2008,Tan2004}
MBAR~\cite{Shirts2008} stores each sample's weights as a matrix $\mathbf{W}$, 
the same matrix of weights discussed earlier in the context of expectations whose entries are $W_{na}$.
The pairwise probabilities can be assembled in what we term the ``overlap matrix,'' constructed from the matrix of weights.
This multidimensional overlap matrix is calculated from the weights as
\begin{equation}
 \mathbf{O} = \mathbf{W}^{\mathbf{T}}\mathbf{W}\mathbf{N} \label{eq:overlap-multi}
\end{equation}
where $\mathbf{N}$ is a diagonal matrix with each $i$th entry equal to the number of samples 
from the $i$th state.~\cite{Klimovich2015,Shirts2008} The individual elements of the overlap 
matrix, $O_{ij}$, 
can be read as the probability of a sample generated in state $j$ being observed in state $i$.
Since $\mathbf{O}$ is a Markov matrix, it can be shown that $\mathbf{O}$ will have at least one eigenvalue of $1$, which is 
also the maximum over all eigenvalues. All other eigenvalues will be real and 
positive.~\cite{Klimovich2015} However, multiple eigenvalues of $1$ in $\mathbf{O}$ indicate that 
there are discontinuous regions of sampled configuration space and $\mathbf{O}$ can be rearranged to a  
block diagonal matrix that illustrates which states are and are not connected. 
Thermodynamic property measurements made between 
the discontinuous configuration spaces will have undefined uncertainty, which numerically 
can show up as either NaN or very large numbers that change dramatically with small changes in sampling.~\cite{Klimovich2015}

Monitoring both the eigenvalues of the overlap matrix and maximum uncertainty can provide good guidance as to
when converged property estimates have been reached. Monitoring exclusively the eigenvalues of 
$\mathbf{O}$ is insufficient to determine convergence over the entire parameter space since $\mathbf{O}$ 
only involves the sampled states.  However, if the sampled states are well-enough dispersed such that the 
estimated uncertainties of all the unsampled states are low, and simultaneously 
all the sampled states are connected as demonstrated by having a single eigenvalue with value 1 for 
$\mathbf{O}$, we can have high confidence that the uncertainty estimates are reliable.

We therefore defined our property estimates as converged  
once there were no repeated eigenvalues of $1$ in $\mathbf{O}$, and once no further clusters of uncertainties in the relative free energy above the target threshold are found, 
i.e. the clustering algorithm can not find new points adjacent grid points with large uncertainty. 
If desired, the uncertainty can be iteratively reduced by lowering the error threshold of the 
algorithm. 
The deviation of the five largest eigenvalues from 1 are shown for each iteration in 
the supplementary material in 
% Table~\ref{tbl:RawIterations}. %xr
Table~S.2.

The proposed adaptive sampling algorithm is only an initial proposal, and sampling 
regions of poor configuration space overlap could be improved by changes to this algorithm. For example, the algorithm discussed 
here and detailed in the supplementary information~\cite{NadenSupNote} identified 
clusters of high relative uncertainty by counting the number of grid points in the 
cluster. This resulted in many states being chosen adaptive at larger $\sii$ due to the 
density of grid points, despite the fact that most of the regions with no phase space 
overlap were at $\sii < 0.35\,\text{nm}$. Although states were eventually placed 
at small $\sii$, one improvement could be to place points in regions with the largest 
integrated uncertainty, using the uncertainty as a weighting on the overall number of grid points. 
This improvement would still favor the larger clusters, but to a lesser extent as new pockets of 
poor configuration space overlap are identified and the uncertainty jumps back up as in 
Fig.~\ref{fig:mean_max}. 

The current study was used to determine high accuracy free energies over entire large parameter 
range, and this study's range of nonbonded parameters will likely exceed many practical applications. 
However, this parameter search method could be adaptively shrunk to hone in on specific property 
estimates.
Instead of determining the thermodynamic properties for a large set of starting parameters,
a desired thermodynamic property could be provided and a set of parameters 
which generate this property are searched for, as is the case for reverse property prediction.
In this case, the initial grid spacing could be larger, and a rough estimate of the 
property surface can be acquired, spending less simulation time per iteration. Each subsequent iteration 
would then narrow the search area and reduce the grid spacing, seeking the target value. 
States from previous iterations outside the narrowed search space, can 
still be included in the analysis using, preventing discarded information. 
Alternatively, computational time can be saved by excluding these outlier states if analysis of 
$\mathbf{O}$ shows that these states are not actually connected to the states ultimately of interest.

Thermodynamic property estimates are not limited to relative solvation free energies, 
entropies, and enthalpies.
Once the simulations are converged, further thermodynamic properties can be derived from 
derivatives and fluctuations with respect to $V$, $P$, and 
$T$,~\cite{Aimoli2014,Avendano2011,Allen1987} as well as any other property computed 
from statistical expectation values.~\cite{Shirts2008}

%%%%%%%%%%%%%%%%%%%%%%%%%%%%%%%%%%%%%%%%%%%%%%%%%%%%%%%%%%%%%%%%%%%%%%%%%%%%%%%%%%%%%%%%%%
\section{Conclusion}

We have shown how one can rapidly estimate thermodynamic properties in a multidimensional 
nonbonded parameter space by combining two time saving advantages. Computing the energies 
required for estimating thermodynamic properties can be accelerated with linear combinations of basis functions 
instead of re-running simulation force loops. Estimating the thermodynamic properties in 
the multidimensional parameter space is possible with the binless, multidimensional, path-free 
statistical method, MBAR. With these methods, properties are estimated at all states of 
interest simultaneously without the need define how any state is connected to any others beforehand. 

Converged results can be acquired by adaptively sampling the multidimensional parameter space, creating 
a network of globally-connected configuration space overlap between all states. Simply adding samples in 
regions of large uncertainty does not necessarily create configuration space overlap to all states, as the 
uncertainty of differences to other states is not reduced. 
Regions of poor configuration space overlap can be identified by examining the overlap 
matrix between all sampled states, and the maximum uncertainty in the unsampled states.
The parameter space is then adaptively sampled until configuration space overlap 
is created between the reference state and all other states of interest, and uncertainties 
are pushed sufficiently low for the purpose at hand.

The methods shown here can help speed up future thermodynamic property searches in 
multidimensional parameter space. So long as the energy functions can be computed 
with vector operations, and do not require re-running the simulation force loops, 
these methods can scale to even higher dimensionality and extend to other thermodynamic properties.
Re-writing the simulation code to directly provide the required basis functions would 
allow even faster energy evaluation, potentially removing the need to ever compute the 
energy of a configuration more than once.

%%%%%%%%%%%%%%%%%%%%%%%%%%%%%%%%%%%%%%%%%%%%%%%%%%%%%%%%%%%%%%%%%%%%%%%%%%%%%%%%%%%%%%%%%%
%%%%%%%%%%%%%%%%%%%%%%%%%%%%%%%%%%%%%%%%%%%%%%%%%%%%%%%%%%%%%%%%%%%%%%%%%%%%%%%%%%%%%%%%%%
\section{Acknowledgments}
The authors would like to thank Kyle Beauchamp and John Chodera at Memorial Sloan-Kettering Cancer Center 
for feedback and assistance in running optimized MBAR code. The authors also acknowledge the support of NSF 
grant CHE-1152786.

%%%%%%%%%%%%%%%%%%%%%%%%%%%%%%%%%%%%%%%%%%%%%%%%%%%%%%%%%%%%%%%%
%BIBLIOGRAPHY information
%%%%%%%%%%%%%%%%%%%%%%%%%%%%%%%%%%%%%%%%%%%%%%%%%%%%%%%%%%%%%%%%
\bibliography{NadenShirtsBasis3}
\end{document}